# Integrating high-quality dielectrics with one-nanometer equivalent oxide thickness on two-dimensional electronic devices

Weisheng Li[1,†], Jian Zhou[1,†], Songhua Cai[2,†], Zhihao Yu[1], Jialin Zhang[3], Nan Fang[4], Taotao Li[2], Yun Wu[5], Tangsheng Chen[5], Xiaoyu Xie[6], Haibo Ma[6], Ke Yan[1], Ningxuan Dai[1], Xiangjin Wu[1], Huijuan Zhao[1], Zixuan Wang[1], Daowei He[1,7], Lijia Pan[1], Yi Shi[1], Peng Wang[2,*], Wei Chen[3], Kosuke Nagashio[4], Xiangfeng Duan[7] & Xinran Wang[1,*]

[1]*National Laboratory of Solid State Microstructures, School of Electronic Science and Engineering and Collaborative Innovation Center of Advanced Microstructures, Nanjing University, Nanjing 210093, China*

[2]*National Laboratory of Solid State Microstructures, Jiangsu Key Laboratory of Artificial Functional Materials, College of Engineering and Applied Sciences and Collaborative Innovation Center of Advanced Microstructures, Nanjing University, Nanjing 210093, China*

[3]*Department of Chemistry, National University of Singapore, 3 Science Drive 3, 117543, Singapore*

[4]*Department of Materials Engineering, The University of Tokyo, Tokyo 113-8656, Japan*

[5]*Science and Technology on Monolithic Integrated Circuits and Modules Laboratory, Nanjing Electronic Device Institute, Nanjing 210016, China*

[6]*School of Chemistry and Chemical Engineering, Nanjing University, Nanjing 210023, China*

[7]*Department of Chemistry and Biochemistry, University of California, Los Angeles, CA, USA.*

† These authors contribute equally to this work.

* Corresponding authors: xrwang@nju.edu.cn, wangpeng@nju.edu.cn

**Two-dimensional (2D) semiconductors are widely recognized as attractive channel materials for low-power electronics. However, an unresolved challenge is the integration of high-quality, ultrathin high-κ dielectrics that fully meet the roadmap requirements for low-power applications. With a dangling-bond free surface, the deposition of dielectrics by atomic layer deposition (ALD) on 2D materials is usually characterized with non-uniform nucleation and island formation, producing a highly porous dielectric layer with serious leakage particularly at the small equivalent oxide thickness (EOT) limit. Here, we report the robust ALD of highly uniform high-κ dielectric on 2D semiconductors by using**

**~0.3 nm-thick exclusively monolayer molecular crystal as seeding layer. Ultrathin dielectrics down to 1 nm EOT is realized on graphene, MoS$_2$ and WSe$_2$, with considerably reduced roughness, density of interface states, leakage current and improved breakdown field compared to prior methods. Taking advantage of the reduced EOT, we demonstrate graphene RF transistors operating at 60 GHz, as well as MoS$_2$ and WSe$_2$ complementary metal-oxide-semiconductor (CMOS) transistors with $V_{dd}$ =0.8 V and ideal subthreshold swing (SS) of 60 mV/dec, 20 nm-channel-length MoS$_2$ transistors with on/off ratio over $10^7$. These studies highlight that our dielectric integration method is generally applicable for different 2D materials, and compatible with top-down fabrication process on large-area chemical vapor deposited films.**

The ever-growing demand for portable electronics requires continuous development of energy-efficient logic devices. In Si CMOS technology, scaling under constrained thermal budget is made possible by the continuous reduction of EOT and the supply voltage ($V_{dd}$) on single transistor level[1]. Currently in the most advanced Si metal-oxide-semiconductor field-effect transistors (MOSFETs) (e.g. Intel 14 nm FinFET structure), the physical thickness of HfO$_2$ is 2.6 nm, corresponding to 0.9 nm EOT[2]. In addition, the gate leakage and interface state density ($D_{it}$) required for low-standby-power CMOS[3] is 1.5×10$^{-2}$ A/cm$^2$ and ~10$^{10}$ cm$^{-2}$eV$^{-1}$. These numbers serve as important benchmarks for any emerging technologies.

To further mitigate SCE in ultra-scaled MOSFET, an alternative route is to reduce the channel thickness, as the characteristic length to avoid SCE scales with $\sqrt{\frac{\varepsilon_{ch}}{N\varepsilon_{ox}}t_{ch}t_{ox}}$. Here, $t_{ch}$ ($\varepsilon_{ch}$) and $t_{ox}$ ($\varepsilon_{ox}$) are thickness (dielectric constant) of channel and gate oxide, $N$ is the number of gate[4]. Therefore, 2D semiconductors, in particular large bandgap transition-metal dichalcogenides (TMDs), have been extensively studied as promising channel materials for low-power electronics[4-7]. However, to integrate high-κ gate dielectric on 2D materials, with sub-1 nm EOT and equivalent leakage current and interface quality as Si CMOS, represents a standing challenge, largely due to difficulties in uniformly nucleates on the intrinsically dangling-bond-free surface of most 2D materials. The ALD deposition on 2D materials was usually achieved through chance nucleation at defects, edges and impurities to produce highly porous thin films[4-7]. Over the past decade, many interfacial activation layers/processes have been developed for uniform deposition of high-κ oxides on 2D materials[8], such as oxidized metal layer[9], organic molecules[10-13], BN[14], ozone[15], mild plasma treatment[16], and electron beam irradiation[17]. However, all these methods have their own drawbacks. For examples, the widely used metal oxidation process suffer from inherent roughness of evaporated metal films as well as damages by high-energy metal ions[18]. Ozone/plasma/electron irradiation processes also involve high-energy and reactive species which can introduce defects and interface states. For molecular layers, it is still difficult to achieve defect-free and precise thickness control over the entire device area[19]. Although exfoliated BN has been proven as an ideal dielectric for 2D materials[20], the leakage current is much higher than the requirement by International Technology Roadmap for Semiconductors (ITRS) due to its low dielectric constant[20], and the aligned transfer process is unscalable. So far, the smallest EOT reported on graphene and TMDs is 1.3 nm and 2.5 nm, respectively[17, 21]. The trade-off between EOT, interface quality and scalability has not been resolved. A transfer gate process has also been explored for integrating high quality gate dielectrics on 2D materials, although it usually involve unusual soft-lithography process and is not readily compatible with current semiconductor manufacturing[22-24].

For low-power logic applications, the requirements of the interfacial buffer layer on 2D materials are extremely stringent. First, the thickness has to approach single atomic layer because any additional thickness would reduce the gate capacitance and make ~1 nm EOT almost impossible. Second, it has to be closely packed on nanometer scale and uniform on micrometer scale to avoid pinholes in the gate dielectric. Third, the interaction should be non-covalent to preserve the intrinsic properties of 2D materials and interfaces, preferably to the level of BN/2D. Finally, the process needs to be robust and scalable to large-area chemical vapor deposition (CVD) films. In this work, we develop a technique to integrate ultrathin high-κ dielectric on graphene, BN and TMDs that satisfies the above requirements simultaneously. We use 3,4,9,10-perylene-tetracarboxylic dianhydride (PTCDA) molecular crystal non-covalently bound to 2D materials as the seeding layer for ALD. The thickness can be controlled precisely to monolayer (ML, ~0.3 nm) using self-limited epitaxy[25], which is a major advancement over prior works[12, 19]. We achieve record-low EOT=1 nm ($t_{ox}$=1.45 nm) on graphene and $MoS_2$ without sacrificing the dielectric properties. Notably, the PTCDA/$HfO_2$ gate dielectric show low leakage current $J < 10^{-2}$ A/cm$^2$ and high breakdown field $E_{bd}$=16.5 MV/cm at 1 nm EOT, which are considerably better than other approaches on 2D materials and meet the ITRS requirement for low-power devices. We also achieve the lowest $D_{it}$ among top-gate $MoS_2$ FETs by ALD. Owing to the low EOT and $D_{it}$, we are able to reduce the $V_{dd}$ of TMD CMOS to the state-of-the-art Si CMOS, with on/off >10$^6$ and nearly ideal switching behavior.

We further demonstrate 20 nm-short-channel MoS$_2$ transistors with on/off >10$^7$ and SS=73 mV/dec. Finally, we grow ML PTCDA/HfO$_2$ on large-area CVD MoS$_2$ films and fabricate top-down transistor arrays with excellent yield and reproducibility. Our technology truly enables the potential of 2D materials for low-power logic applications.

**Deposition of high-κ dielectrics on 2D materials using ML PTCDA as seeding layer**

Fig. 1a illustrates the HfO$_2$/ML PTCDA/2D material hybrid structure studied in this work. PTCDA is well known to have a layered crystal structure, with 0.3 nm interlayer distance corresponding to the π-π stack direction (Supplementary Fig. S1). When grown on 2D surfaces like graphene, it self-assembles into a herringbone lattice stabilized by hydrogen bonding[26]. To obtain ultrathin EOT, controllable growth of exclusively ML PTCDA seeding layer is crucial. We carried out self-limited van der Waals epitaxial growth in a tube furnace described by our previous work[25]. The ML PTCDA film was highly crystalline and uniform as confirmed by atomic force microscopy (AFM) (Fig. 1c, Supplementary Fig. S1), cross-polarized optical microscopy (Supplementary Fig. S2) and scanning tunneling microscopy (STM) (Fig. 1a inset). High-resolution STM clearly revealed herringbone packing, consistent with previous experiments[26] and theoretical calculations[27]. The self-limited growth could be understood by the thickness-dependent binding energy calculated by molecular dynamics simulations (Fig. 1b). For graphene and BN, the binding energy on substrate is higher than that on additional PTCDA layers due to strong π-π interaction, indicating that the first layer is thermodynamically more stable. Although the trend was opposite on MoS$_2$, we were still able to grow uniform ML as shown by the height increase of 0.3-0.4 nm in AFM (Fig. 1d, Supplementary Fig. S1) and scanning transmission electron microscopy (STEM) (Fig. 2).

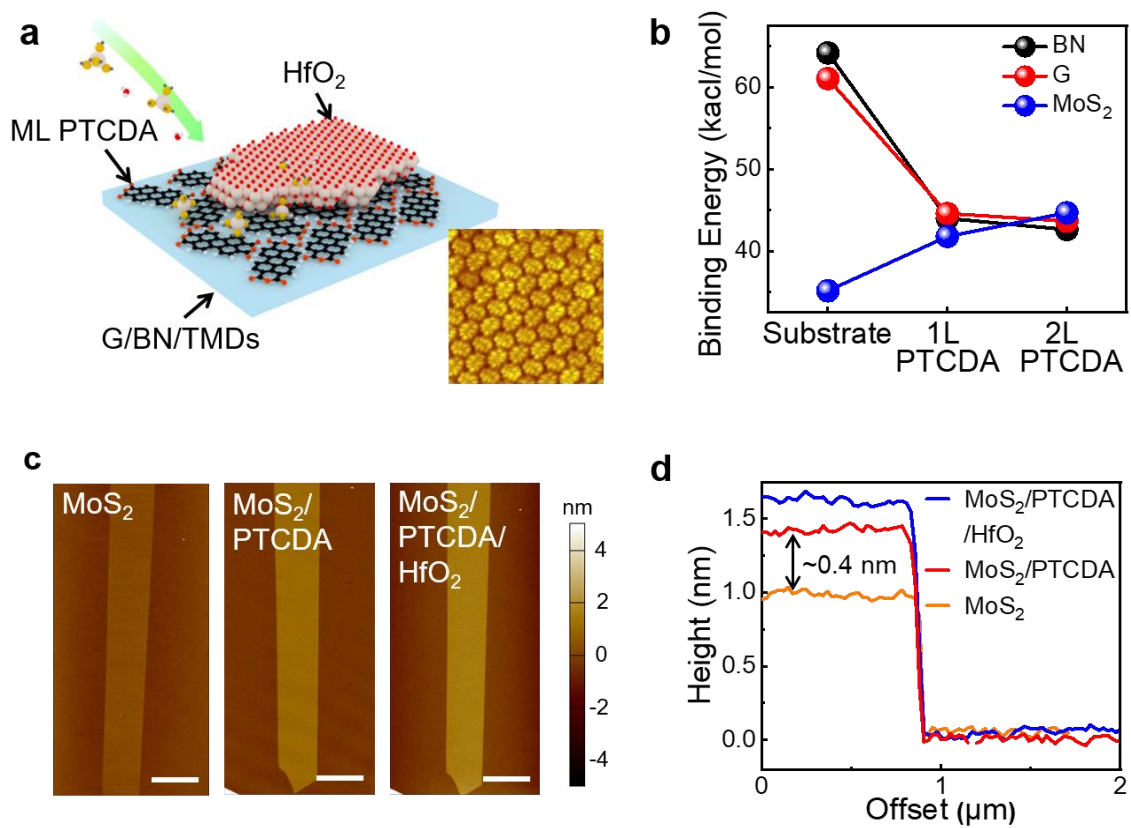

Figure 1. **Deposition of ultrathin high-k oxide on 2D materials. a.** Schematic illustration of the hybrid ML PTCDA/HfO$_2$ gate stack on 2D materials. Inset shows a 10 nm×10 nm high-resolution STM scan of ML PTCDA on graphene. **b.** Calculated PTCDA-substrate binding energy on different 2D materials. **c.** AFM images of an exfoliated MoS$_2$ (left), after ML PTCDA deposition (middle) and after 2 nm HfO$_2$ deposition (right). Scale bars are 3 μm. **d.** Height profile measured from **c**.

Following adsorption on 2D materials, the carbonyl functional groups in PTCDA molecules serve as closely distributed sites for ALD nucleation. Indeed, graphene became more hydrophilic after ML PTCDA deposition, with water contact angle decreased from 95 ° to 62 ° (Supplementary Fig. S3). The large binding energy stabilized the PTCDA layer at our ALD temperature of 150 °C as confirmed by Raman spectroscopy (Supplementary Fig. S4). As a result, we were able to deposit pinhole-free HfO$_2$ on graphene, BN and TMDs (including MoS$_2$, WS$_2$, MoSe$_2$ and WSe$_2$) with extremely low mean-square roughness of 0.13 nm (Fig.

1c, d, Supplementary Fig. S1). The roughness is the lowest on 2D materials so far (Table S1), proving that ML PTCDA crystal is an excellent seeding layer. As expected, the control samples without PTCDA showed discontinuous oxide deposition (Supplementary Fig. S1).

The pristine quality of the dielectric interface was further revealed by STEM (Fig. 2, Supplementary Fig. S5). Fig. 2a-c are bright-field cross-section STEM images of ML PTCDA/$HfO_2$ dielectric stack on graphene, BN and $MoS_2$, respectively, showing very smooth $HfO_2$ down to 2 nm. It is worth mentioning that the same uniformity was maintained over the entire device area of several micrometers consistent with AFM (see the white horizontal line in the low magnification image in Supplementary Fig. S5a). The ML PTCDA was clearly observed as an additional layer above graphene and BN (Fig. 2a, b), with an interlayer distance of ~0.31 nm corresponding to the van der Waals gap. The PTCDA layer was rougher than the topmost layer of 2D materials, possibly due to the damage by focus ion beam during sample preparation. The PTCDA layer appeared as a bright straight line on $MoS_2$ because Mo and S elements are much heavier than C (Fig. 2c). To verify the elemental distribution and existence of PTCDA on $MoS_2$, STEM-EELS mapping was carried out in this region. By extracting the signals of S, O and C edges in EELS spectrum from every scanning pixel, the distribution of these elements was mapped out (Fig. 2d), and corresponding well to the device structure. We clearly observed a high concentration of C signal at the $HfO_2$/$MoS_2$ interface. Close inspection of the EELS spectrum revealed that the main C peak was red-shifted compared to amorphous C contaminations from other positions (Fig. 2e), due to the existence of C=O bond in PTCDA[28]. The same trend was followed in the case of PTCDA on BN (Fig. S5d), proving unequivocally that ML PTCDA served as the seeding layer. To compare with other growth methods, we also performed STEM characterizations on a control sample with 1 nm evaporated Al buffer layer and 4 nm $HfO_2$ (Supplementary Fig. S5c), which shows damage of 2D material as well as fluctuation of oxide thickness by 0.6 nm. The quality of stack is clearly inferior to our PTCDA/$HfO_2$ stack.

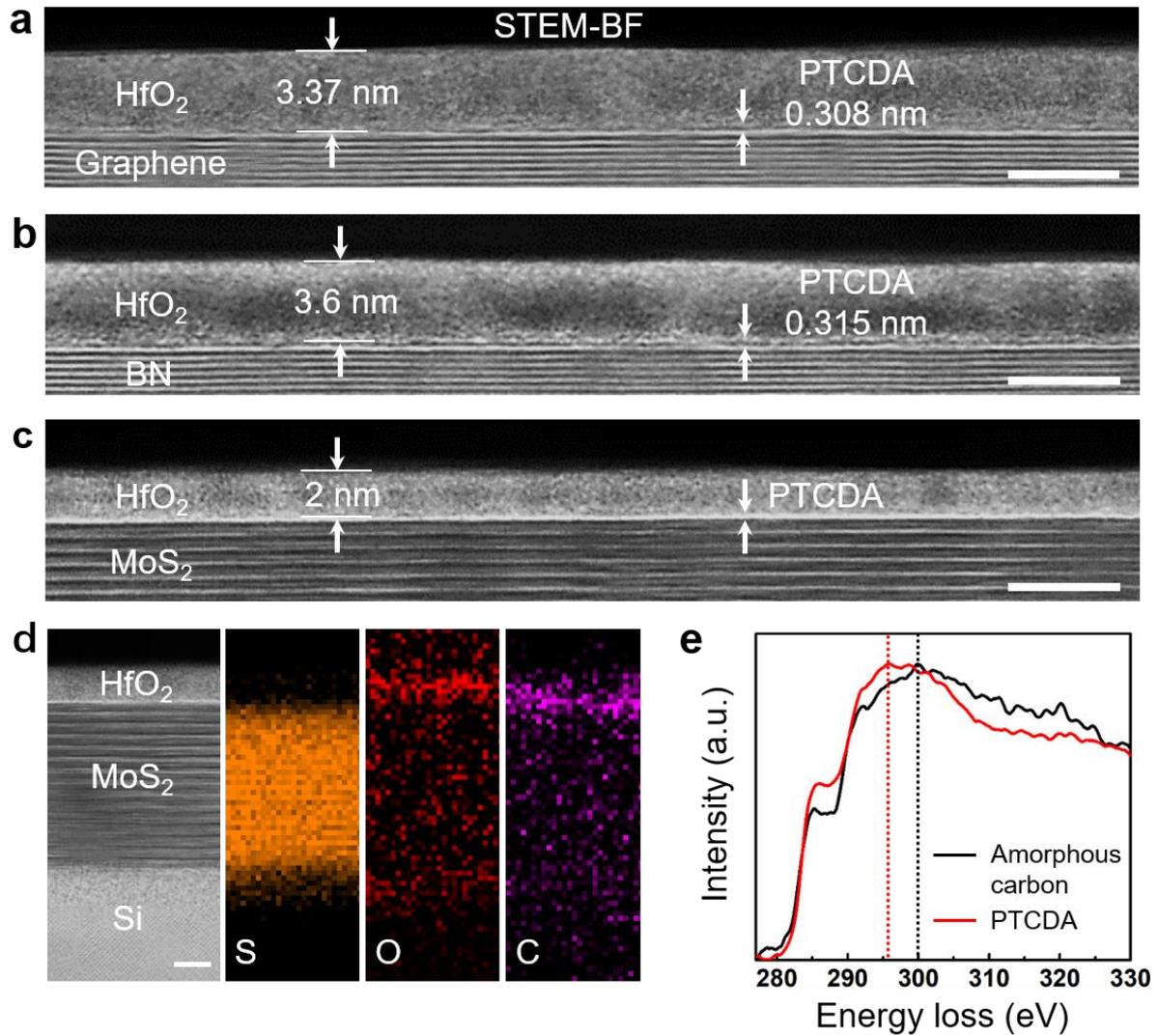

Figure 2. **STEM characterization of gate stacks. a-c.** Cross-sectional STEM Bright-field image of ML PTCDA/HfO$_2$ on graphene (**a**), MoS$_2$ (**b**) and h-BN (**c**), respectively. Scale bar are 5 nm. The thickness of HfO$_2$ and PTCDA layer is marked, respectively. **d.** STEM-EELS mapping. From left to right, STEM bright-field image and corresponding elemental mapping of S, O and C respectively. Scale bar is 2.5 nm **e.** EELS spectra of carbon *K*-edge signal from PTCDA at the HfO$_2$/MoS$_2$ interface (the high intensity area of C in **d**) and amorphous carbon from other positions.

**Dielectric properties on graphene FETs**

To study the dielectric properties, we fabricated double-gate graphene FET structure with 275 nm SiO$_2$/Si as the global backgate and ML PTCDA/HfO$_2$/Au as the top-gate stack (Fig. 3a inset). Double-sweep $I_D$-$V_{TG}$ transfer characteristics showed mobility $\mu$~3500 cm$^2$/Vs, no hysteresis and residual carrier density on the order of 10$^{11}$ cm$^{-2}$ (Fig. 3a, Supplementary Fig. S6), typical for graphene FET on SiO$_2$[29]. The fabrication of top-gate did not introduce defects, doping or interface states as corroborated by the step-by-step electrical measurements (Supplementary Fig. S6). To accurately extract topgate capacitance ($C_{TG}$) and EOT, we used two independent methods, namely double-gate measurement on graphene FET (Fig. 2b, Supplementary Fig. S7) and capacitance measurement (C-V and C-f) (Supplementary Fig. S8). Capacitance measurement was performed on graphite (MoS$_2$)/ML PTCDA/HfO$_2$/Au capacitors on quartz substrate to eliminate parasitic capacitance. Theoretically, $C_{TG}$ comprises of contributions from both PTCDA and HfO$_2$:

$$\frac{1}{C_{TG}} = \frac{t_{PTCDA}}{\varepsilon_0 \varepsilon_{PTCDA}} + \frac{t_{ox}}{\varepsilon_0 \varepsilon_{ox}} \qquad (1)$$

where $t_{PTCDA}$ ($t_{ox}$) and $\varepsilon_{PTCDA}$ ($\varepsilon_{ox}$) is physical thickness and dielectric constant of PTCDA (HfO$_2$), and $\varepsilon_0$ is vacuum dielectric constant. EOT is calculated as

$$EOT = \frac{3.45\ \mu F/cm^2}{C_{TG}} \qquad (2)$$

where 3.45 $\mu F/cm^2$ is the gate capacitance for 1 nm SiO$_2$. The effective dielectric constant of the whole stack (PTCDA+HfO$_2$) is calculated as

$$\varepsilon_{eff} = \frac{C_{TG}(t_{PTCDA}+t_{ox})}{\varepsilon_0} \qquad (3)$$

Fig. 3c summarizes the EOT and $\varepsilon_{eff}$ as a function of $t_{ox}$ measured from 30 graphene FETs, 35 MoS$_2$ FETs and 6 capacitors (4 graphene and 2 MoS$_2$). The four sets of data give nearly identical results, showing excellent reproducibility on different 2D materials. The EOT scales linearly with $t_{ox}$ and intercepts the Y axis at 0.6 nm, which corresponds to the EOT of ML PTCDA. Both EOT and $\varepsilon_{eff}$ are well described by theory (see the fittings in Fig. 3c) using $\varepsilon_{PTCDA}$=1.9 and $\varepsilon_{ox}$=16.8 consistent with literatures[19, 30]. Remarkably, we fabricated several working devices with $t_{ox}$=1.45 nm and EOT=1 nm using only 7 cycles of ALD ($C_{TG}/C_{BG}$=276, Fig. 3b). Both $t_{ox}$ and EOT are record-low on topgate 2D FETs (Table S1, S2). This further proves that the ML PTCDA layer is defect-free across the entire channel region. For comparison, ALD seeded by evaporated PTCDA on graphene/SiC only achieved lowest $t_{ox}$=10 nm and EOT=3.87 nm without gate failure[19].

Organic dielectrics generally suffer from low operating frequency[31]. However, the capacitance of PTCDA/HfO$_2$ maintained nearly constant up to 1 MHz (Supplementary Fig. S8).

To test whether the gate dielectric functions at higher frequency, we fabricated ground-signal-ground graphene radio-frequency (RF) transistors on highly resistive Si substrate (Fig. 3e inset), with channel length of 500 nm, gate length ($L_g$) of 360 nm and $t_{ox}$=6 nm (EOT=2 nm). The device showed $\mu$~2000 cm$^2$/Vs and transconductance of 48 $\mu$S/$\mu$m under $V_D$=0.05 V (Supplementary Fig. S9). Owing to the ultrathin EOT, the transconductance (scaled by $V_D$) was nearly one order of magnitude higher than similar device with $L_g$=550 nm and comparable to the self-aligned structure with $L_g$=300 nm[32, 33]. Fig. 3e plots the small-signal current gain $|h_{21}|$ extracted from the measured S parameters from 60 MHz to 20 GHz using an Agilent N5247A network analyzer. The $|h_{21}|$ showed a typical *1/f* frequency dependence expected for an ideal FET, and the cut-off frequency ($f_T$) of 10.9 GHz was obtained from the linear fit. We further performed careful de-embedding using the exact pad layout as "open," "short," and "through" structures on the same chip[32]. The intrinsic $f_T$ reached 60 GHz, which outperformed previously reported graphene RF transistors (under the same $L_g$) by more than 100%[33]. We conclude that the PTCDA layer does not degrade the dielectric properties up to tens of GHz.

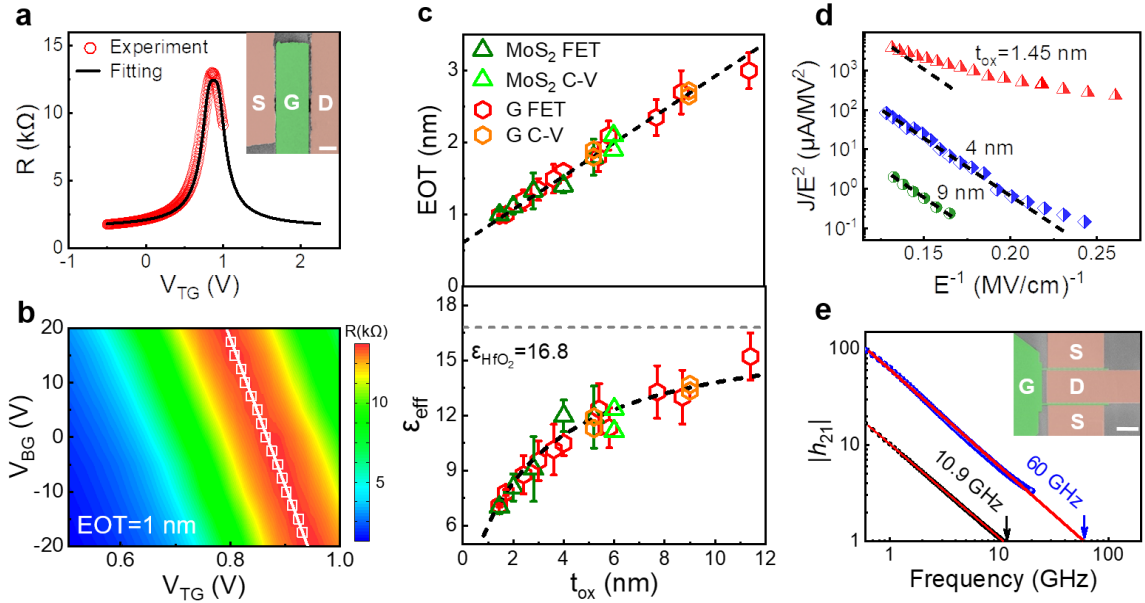

Figure 3. **Dielectric properties on graphene FETs. a.** Double-sweep resistance-$V_{TG}$ characteristics of a topgate graphene FET with 7 cycles of ALD HfO$_2$ (symbols) and theoretical fitting (line). Inset shows false color SEM image of the device structure. Scale bar is 2 μm. **b.** Resistance as a function of $V_{TG}$ and $V_{BG}$ of the same device in **a**. The Dirac points are shown by the white symbols, and linear fitting yields EOT=1 nm. **c.** EOT (top panel) and $\varepsilon_{eff}$ (bottom panel) as a function of $t_{ox}$. Red (orange) hexagon and green (bright green) triangle

symbols represent results from graphene FET (capacitor) and MoS$_2$ FET (capacitor) measurements, respectively. Dashed lines are theoretical fittings. The bar graph shows the mean values (n = 30 for graphene FETs, n=35 for MoS$_2$ FETs), and error bars are the standard deviation. **d.** F-N plot of leakage current for $t_{ox}$=9 nm (green), 4 nm (blue), and 1.45 nm (red). The dashed lines are linear relationship as expected for F-N tunneling. **e.** Small-signal current gain $|h_{21}|$ measured from a graphene FET with channel length of 500 nm and $L_g$=360 nm before (black symbol) and after (blue symbol) de-embedding. Measurement conditions: $V_D$=1 V, $V_{TG}$=-0.4 V. Red lines are linear fittings to extract $f_T$. Inset shows false color SEM image of the device. Scale bar is 4 μm.

As $t_{ox}$ reduces, gate leakage from quantum mechanical tunneling becomes a major source of power consumption in integrated circuits. We investigated the tunneling mechanism for a wide range of $t_{ox}$. For $t_{ox}$=9 nm, the leakage current could be well described by Fowler-Nordheim (F-N) tunneling through a triangular potential barrier, $J \sim E_{ox}^2 \exp(-B/E_{ox})$, where $E_{ox}$ is electric field in the oxide (see F-N plot in Fig. 3d). For $t_{ox}$=4 nm, $J$ deviated from straight line in F-N plot at low field, suggesting that direct tunneling started to become significant. For $t_{ox}$=1.45 nm (EOT=1 nm), direct tunneling became dominant and led to exponential increase of leakage and complete deviation from F-N fitting. Nevertheless, the leakage current was reversible and dielectric did not breakdown until $V_{TG}$=3.27 V (Supplementary Fig. S10), suggesting excellent uniformity even down to such extreme thickness. We emphasize that the insertion of ML PTCDA significantly reduces the gate leakage and increases the breakdown field. Quantitative assessment and benchmark with other technologies will be discussed in the last section.

**Low-power TMD CMOS**

After characterizing the dielectric properties, we next integrated ML PTCDA/HfO$_2$ on TMDs for high on/off ratio CMOS. MoS$_2$ and WSe$_2$ were selected as representative channel materials for NMOS and PMOS, respectively. The device structure was similar to graphene FET (Fig. 4a), but with the backgate oxide changed to 30 nm Al$_2$O$_3$ to screen the Coulomb impurities[34]. Fig. 4b and c present transfer and output characteristics of ML MoS$_2$ and WSe$_2$ FETs with $t_{ox}$=3 nm (EOT=1.3 nm, obtained from double-gate measurement).The NMOS (PMOS) exhibits on/off ratio of $10^7$ ($10^6$) and near-Boltzmann-limit SS of 60 (67) mV/dec. The SS is maintained for several orders of $I_D$ for both sweeping directions (Supplementary Fig. S11b). The output characteristics are linear at low $V_D$, indicating Ohmic contacts. Importantly, $I_D$ already shows saturation at $V_D$=0.8 V because of channel pinch-off enforced by small EOT. This allows us to reduce the operation voltage of TMD-based CMOS to state-of-the-art Si CMOS for the first time[11, 15, 16, 30, 35]. The ultrathin EOT further allows us to realize a 2D CMOS inverter with gain of over 12 V/V at $V_{dd}$=0.5 V (Fig. 4c and Supplementary Fig. S12). Benefit from the low $V_{dd}$, the static and dynamic power consumption of the inverter is less than 0.6 nW and 0.8 nW, respectively, superior to any reported 2D inverters[36]. Therefore, our dielectric integration technology truly enables the benefit of 2D channel in reducing the power consumption of integrated circuit. Like graphene FET, we also achieved the lowest EOT=1 nm on $MoS_2$ NOMS with SS = 61 mV/dec (Supplementary Fig. S11). To further evaluate the process reproducibility, we measured 20 $MoS_2$ (1-3 layer) NMOS with $t_{ox}$=3 nm. The average two-terminal mobility, SS and EOT was 34 $cm^2$/Vs, 66 mV/dec and 1.35 nm, respectively (Fig. 3e, f). Both SS and EOT showed narrow Gaussian distribution with less than 15% variation (measured by the full-width-of-half-maximum from the Gaussian fit). On the other hand, control devices with 3 nm neat $HfO_2$ topgate dielectric shows serious leakage due to the non-uniform nucleation.

The 60mV/dec switching, hysteresis-free transfer characteristics (Supplementary Fig. S13) and small frequency dispersion of $C$-$V$ curves (Supplementary Fig. S8) consistently suggest very low $D_{it}$. To quantitatively analyze the interface properties, we performed modeling of $I_D$-$V_{TG}$ characteristics by considering the $MoS_2$ channel carrier statistics through quantum capacitance and transport through Drude model[37]. To avoid the underestimation of $D_{it}$, we adopted a conservative approach and analyzed ML $MoS_2$ FETs with relatively large EOT=2.1 nm. Fig. 3g shows the extracted $D_{it}$ distribution. At the mid-gap region corresponding to the small-current subthreshold region with constant SS, $D_{it}$ is as low as ~ $8\times10^{11}$ $cm^{-2}eV^{-1}$ and gradually increases to the range of $10^{12}$ $cm^{-2}eV^{-1}$ toward the conduction band edge. Compared with previous metal-buffered ALD $MoS_2$ FETs[37] and other top-gate $MoS_2$ FETs[15, 30, 38-41], our device shows the lowest $D_{it}$ at the mid-gap region, which indicates that the ideal SS is not only due to the ultrathin EOT but also due to the improved interface properties. The origin for this improvement is attributed to the non-invasive and atomically flat nature of the PTCDA buffer layer, low strain due to van der Waals separation as well as indirect ALD deposition. The dense and pinhole-free oxide also well encapsulates the $MoS_2$ channel to give excellent long-term stability of the devices (Supplementary Fig. S14).

Moving forward to integration, it is important to demonstrate uniform dielectric deposition on large-area films. To this end, we deposited ML PTCDA/$HfO_2$ on centimeter-scale CVD $MoS_2$ transferred off sapphire substrate. Similar to exfoliated samples, conformal deposition was realized despite unavoidable PMMA residues from the transfer process (Supplementary Fig. S15). We next fabricated topgate transistor arrays using standard top-down lithography and etching processes (Fig. 4h-i, see Supplementary Fig. S16 for the process flow). Even though the process involved multiple steps of lithography, we were still able to achieve device yield of 90% (without gate failure). The transfer characteristics of 27 devices show small variations in threshold voltage and SS (Fig. 4i). The average SS of 160 mV/dec is much larger than exfoliated $MoS_2$ devices, which is attributed to thicker oxide ($t_{ox}$=6 nm) and additional interface states associated with CVD samples. We also developed new transfer technology of CVD TMDs using PTCDA/oxide stack instead of widely used polymers as support layer (see Method and Supplementary Fig. 17). Centimeter-scale continuous CVD $MoS_2$ film on sapphire could be transferred to target substrate with high integrity, extremely clean interface and small roughness. Compared to conventional approaches, our method can avoid any polymer and solvent contaminations on TMDs surface and form high-quality dielectric interface on large-area films. With further process optimization, it can provide a new path for TMD-based large-scale integrated circuits.

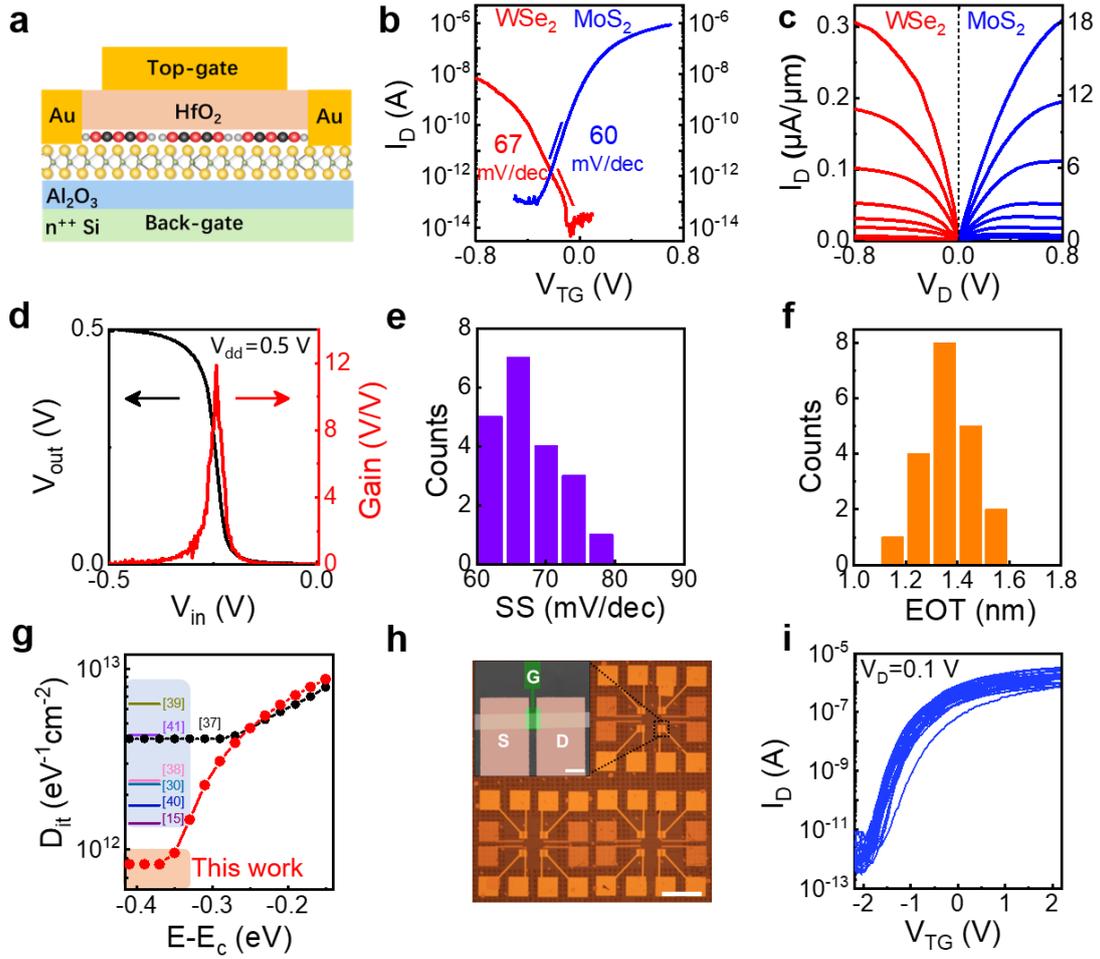

Figure 4. **Low-power TMD CMOS. a.** Schematic of the device structure. **b.** Transfer characteristics of a ML $MoS_2$ NMOS (blue) and $WSe_2$ PMOS (red) with $t_{ox}$=3 nm and EOT=1.3 nm. The SS values are marked on the plot. Measured under $V_D$=0.5 V for both devices, and $V_{BG}$=0 V (-8 V) for $MoS_2$ ($WSe_2$). **c.** Output characteristics of the same devices in **b**. For the $MoS_2$ device, from top to bottom, $V_{TG}$ from 0.5 V to -0.3 V with 0.1 V step, $V_{BG}$=15 V to turn on the ungated region. For the $WSe_2$ device, from top to bottom, $V_{TG}$ from -2 V to -1.2 V with 0.1 V step, $V_{BG}$=-15 V to turn on the ungated region. The $WSe_2$ device shows much lower on-state current density than the $MoS_2$ device due to the crystal quality and contact resistance. **d.** Measured output voltage and gain as a function of input voltage of a CMOS inverter under $V_{dd}$=0.5 V. **e. f.** Statistical distribution of SS (**e**) and EOT (**f**) measured from 20 $MoS_2$ FETs with $t_{ox}$=3 nm. **g.** $D_{it}$ distribution as a function of energy below the conduction band from a typical $MoS_2$ FET with EOT=2.1 nm, in comparison with mid-gap $D_{it}$ values from literatures[15, 30, 37-41]. **h.** Optical micrograph of $MoS_2$ FET arrays fabricated on CVD samples. Scale bar is 150 μm. Inset shows the false color SEM image of a CVD MoS$_2$ device. We adopted overlap structure between topgate and source/drain to minimize the ungated region[42]. Scale bar is 7 μm. **i.** Transfer characteristics of 27 CVD MoS$_2$ FETs.

**Short-channel MoS$_2$ FET**

For Si CMOS, the industry has adopted tri-gate structure since 22 nm node because of the otherwise prohibitive SCE in planar devices[43]. On the other hand, the combination of ultrathin EOT and 2D channel should in principle minimize SCE[5]. To prove this point, we designed and fabricated short-channel ($L_g$~20 nm, limited by our lithographic resolution) MoS$_2$ FETs with graphene backgate and ML PTCDA/HfO$_2$ as gate dielectric (Fig. 5a, b, see Supplementary Fig. S18 for process flow). The MoS$_2$ was dry transferred onto the backgate stack[44]. This structure can fully exploit the combination of ultrathin dielectric grown on 2D materials (in this case graphene) and ultra-scaled channel without complicated topgate alignment. Fig. 5c and d plot the transfer and output characteristics of a typical device with $L_g$=20 nm and EOT=2 nm. The device shows on/off ratio in excess of $10^7$ and steep SS=73 mV/dec, suggesting that the structure is indeed immune to SCE. The SS is lower than most reported MoS$_2$ FETs with similar channel length (Table S2) and is close to 69 mV/dec for 22 nm node Si tri-gate transistors[43]. We note that the current density of our short-channel devices is limited by contact resistance, which will be optimized in future works.

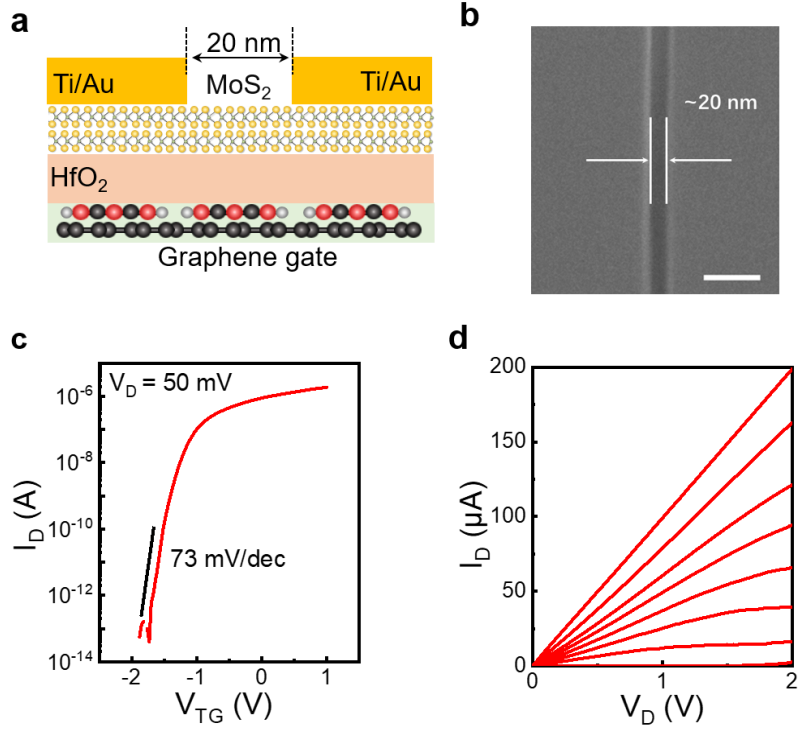

Figure 5. **Short-channel MoS$_2$ FETs. a.** Schematic of the device structure. **b.** SEM image of a device with $L_g$=20 nm. Scale bar is 80 nm. **c.** Transfer characteristics of the device shown in **b**, measured under $V_D$=50 mV. **d.** Output characteristics of the same devices in **c** under $V_{BG}$=0 V. From top to bottom, $V_{TG}$ from -2 V to 2 V with 0.5 V step.

**Benchmark with other technologies**

In order to critically assess the potential for applications, it is important to benchmark with mature Si technology and other approaches on 2D materials. The breakdown characteristics (Fig. 6a) and gate leakage (Fig. 6b) are most important factors for reliability and power consumption. Fig. 6a plots the breakdown voltage ($V_{bd}$) and field ($E_{bd} = \frac{V_{bd}}{t_{PTCDA}+t_{ox}}$) of topgate graphene FETs (Supplementary Fig. S10) as a function of EOT. We find that $V_{bd}$ increases linearly with EOT in the whole range, and a linear fitting yields a constant maximum carrier density $n_{max} = \frac{C_{TG}V_{bd}}{e} = \frac{\varepsilon_{SiO_2}\varepsilon_0}{e}\frac{V_{bd}}{EOT} = 6.5 \times 10^{13}\ cm^{-2}$. This is the highest carrier density achieved in 2D materials using oxide[11, 16, 19, 30, 45], even close to that of ionic liquid gating[46]. On the other hand, $E_{bd}$ shows the opposite trend of decreasing with EOT (and $\varepsilon_{eff}$), which is in accordance with $E_{bd} \sim \varepsilon^{-1/2}$ for high-κ materials[47] (see the blue dashed line in Fig. 6a for theoretical fitting). The $E_{bd}$ is within the theoretical limits predicated for $HfO_2$ (4-8 MV/cm)[47] at large EOT but rapidly increases to 16.5 MV/cm at 1 nm EOT. This indicates that the ML PTCDA actually controls the breakdown at small EOT, which again proves its excellent quality as buffer layer. Under the same EOT, our $V_{bd}$ is significantly higher than most literature results on 2D materials using metal and organic buffer layer[11, 16, 19, 30, 45]. Compared with the best result of 5 nm $Hf/HfO_2$ on graphene (EOT=1.38 nm) reported by Samsung[48], our $V_{bd}$ is similar but $E_{bd}$ is increased by more than 25%.

Fig. 6b compares the gate leakage current of topgate graphene and $MoS_2$ FETs with available data on Si CMOS[2, 49] and 2D materials[11, 14, 16, 19, 30, 45, 48], under the same $V_{TG}$=1 V. Surprisingly, our data points almost overlap with $HfO_2$/Si deposited by advanced industrial tools[49], and are orders of magnitude lower than $SiO_2$/Si and most 2D data. Even at 1 nm EOT, the leakage current satisfies the low-power requirement for CMOS ($1.5 \times 10^{-2}$ A/cm$^2$)[49]. Although BN is considered as an ideal dielectric for 2D materials, the leakage current of graphene/BN/graphene structure at EOT=1 nm is over 5 orders of magnitude higher than our results[49]. Therefore, BN is not suitable for low-power 2D transistors. We note that although the PTCDA layer decreases $\varepsilon_{eff}$ at small EOT, the leakage current does not deviate from standalone $HfO_2$ on Si. This is probably due to the 0.3 nm-wide van der Waals gap between PTCDA and 2D materials, which effectively suppresses the tunnel current.

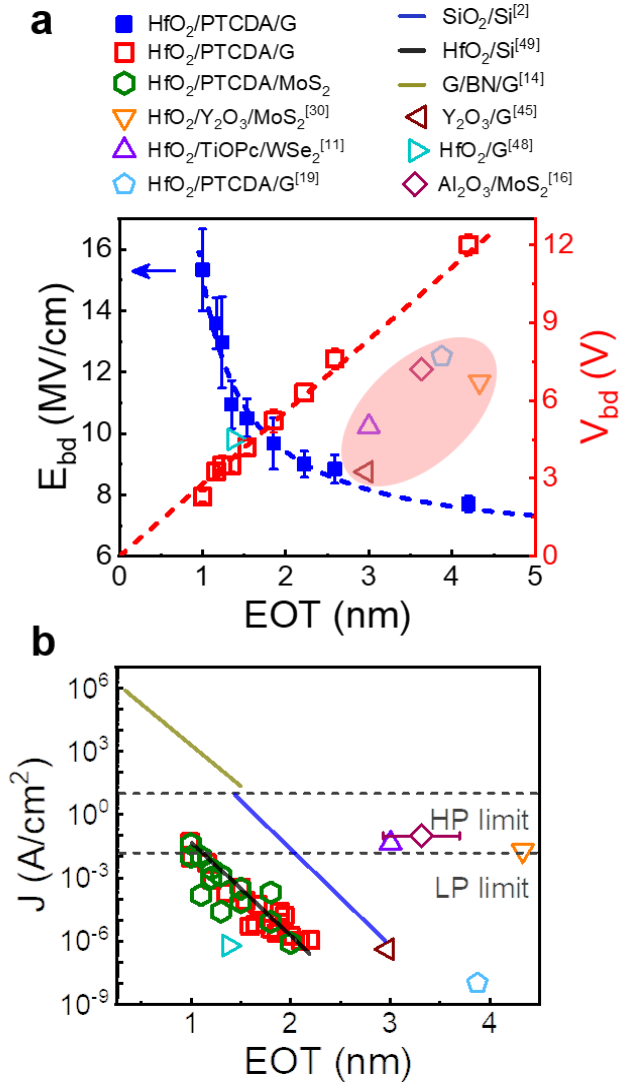

Figure 6. **Benchmark of breakdown and leakage characteristics with other technologies. a.** $E_{bd}$ (closed symbols, left axis) and $V_{bd}$ (open symbols, right axis) as a function of EOT measured from topgate graphene FETs, in comparison with 2D data from literature[11, 16, 19, 30, 45, 48]. The dashed red line is linear fitting of $V_{bd}$ which gives maximum carrier density of $6.5 \times 10^{13}$ cm$^{-2}$. The blue dashed line is the theoretical fitting by $E_{bd} \sim \varepsilon_{eff}^{-1/2}$, where the $\varepsilon_{eff}$ − EOT relationship is obtained from Fig. 3c. The bar graph shows the mean values (n = 30), and error bars are the standard deviation. **b.** Leakage current (under $V_{TG}$=1 V) as a function of EOT measured from topgate graphene (red symbols) and MoS$_2$ (green symbols) FETs, in comparison with Si and 2D data from literature[11, 16, 19, 30, 45, 48]. The black, blue and gold solid lines are leakage current of HfO$_2$/Si[49], SiO$_2$/Si[2] and graphene/BN/graphene[14]. The two

horizontal dashed lines are the ITRS limit for high-performance and low-power logic applications[2].

**Conclusion and outlook**

In conclusion, we achieved record-low EOT=1 nm on graphene and TMDs using self-limited ML molecular crystal as buffer layer. The gate dielectric shows pristine interface quality, low leakage current and high breakdown field simultaneously. For the first time, the leakage current is reliably reduced to the level of Si CMOS and comply with the low-power requirement down to 1 nm EOT. The combined EOT and interface integrity lead to ideal 60 mV/dec switching, low $V_{dd}$=0.8 V, and reduced SCE in TMD transistors. Our process has little variations on different 2D materials, and is applicable to large-area CVD films. Although we exclusively use $HfO_2$ in this work, our method is generally applicable to any oxides that can be deposited by ALD (see Fig. S19 for additional data on $Al_2O_3$). With higher $\varepsilon_{ox}$ oxides such as $ZrO_2$ ($\varepsilon_{ox}$=35)[50], it is possible to achieve EOT=0.77 nm at $t_{ox}$=1.5 nm. Further reduction of EOT would require the substitution of low-κ organic seeding layer (while maintaining the same uniformity), which is very challenging. High-performance integrated circuits will require more efforts toward cleaner CVD growth, transfer and fabrication processes that minimize surface impurities[18].

**Data availability**

The data that support the plots within this paper and other findings of this study are available from the corresponding author upon reasonable request.

**References:**

1.  Theis, T.N. & Wong, H.S.P. The End of Moore's Law: A New Beginning for Information Technology. *Comput. Sci. Eng.* **19**, 41-50 (2017).

2.  Natarajan, S. et al. A 14nm Logic Technology Featuring 2nd-Generation FinFET Interconnects, Self-Aligned Double Patterning and a 0.0588 μm² SRAM cell size. *Inter. Electron Dev. Meet. Tech.*

*Digest* 3.7.1-3.7.3 (2014).

3. Robertson, J. High dielectric constant oxides. *Eur. Phys. J-Appl. Phys.* **28**, 265-291 (2004).

4. Chhowalla, M.D.J.A. Two-dimensional semiconductors for transistors. *Nat. Rev. Mater.* **11**, 16052 (2016).

5. Fiori, G. et al. Electronics based on two-dimensional materials. *Nat. Nanotechnol.* **9**, 768-779 (2014).

6. Desai, S.B. et al. $MoS_2$ transistors with 1-nanometer gate lengths. *Science* **354**, 99-102 (2016).

7. Radisavljevic, B., Radenovic, A., Brivio, J., Giacometti, V. & Kis, A. Single-layer $MoS_2$ transistors. *Nat. Nanotechnol.* **6**, 147-150 (2011).

8. Kim, H.G. & Lee, H. Atomic Layer Deposition on 2D Materials. *Chem. Mater.* **29**, 3809-3826 (2017).

9. Zhu, Y. et al. Monolayer Molybdenum Disulfide Transistors with Single-Atom-Thick Gates. *Nano Lett.* **18**, 3807-3813 (2018).

10. Wang, X., Tabakman, S.M. & Dai, H. Atomic Layer Deposition of Metal Oxides on Pristine and Functionalized Graphene. *J. Am. Chem. Soc.* **130**, 8152-8153 (2008).

11. Park, J.H. et al. Atomic Layer Deposition of $Al_2O_3$ on $WSe_2$ Functionalized by Titanyl Phthalocyanine. *ACS Nano* **10**, 6888-6896 (2016).

12. Alaboson, J.M.P. et al. Seeding Atomic Layer Deposition of High-κ Dielectrics on Epitaxial Graphene with Organic Self-Assembled Monolayers. *ACS Nano* **5**, 5223-5232 (2011).

13. Johns, J.E., Karmel, H.J., Alaboson, J.M.P. & Hersam, M.C. Probing the Structure and Chemistry of Perylenetetracarboxylic Dianhydride on Graphene Before and After Atomic Layer Deposition of Alumina. *J. Phys. Chem. Lett.* **3**, 1974-1979 (2012).

14. Britnell, L. et al. Electron Tunneling through Ultrathin Boron Nitride Crystalline Barriers. *Nano Lett.* **12**, 1707-1710 (2012).

15. Wang, J. et al. Integration of High-k Oxide on $MoS_2$ by Using Ozone Pretreatment for High-Performance $MoS_2$ Top-Gated Transistor with Thickness-Dependent Carrier Scattering

Investigation. *Small* **11**, 5932-5938 (2015).

16. Wang, X. et al. Improved integration of ultra-thin high-k dielectrics in few-layer $MoS_2$ FET by remote forming gas plasma pretreatment. *Appl. Phys. Lett.* **110**, 53110 (2017).

17. Xiao, M., Qiu, C., Zhang, Z. & Peng, L. Atomic-Layer-Deposition Growth of an Ultrathin $HfO_2$ Film on Graphene. *ACS Appl. Mater. Inter.* **9**, 34050-34056 (2017).

18. Liu, Y. et al. Approaching the Schottky–Mott limit in van der Waals metal–semiconductor junctions. *Nature* **557**, 696-700 (2018).

19. Sangwan, V.K. et al. Quantitatively Enhanced Reliability and Uniformity of High-κ Dielectrics on Graphene Enabled by Self-Assembled Seeding Layers. *Nano Lett.* **13**, 1162-1167 (2013).

20. Dean, C.R. et al. Boron nitride substrates for high-quality graphene electronics. *Nat. Nanotechnol.* **5**, 722-726 (2010).

21. Yang, L., et al. 10 nm nominal channel length $MoS_2$ FETs with EOT 2.5 nm and 0.52 mA/µm drain current. *73rd Annual Device Research Conference*. IEEE, 237-238(2015).

22. Liao, L. et al. Top-Gated Graphene Nanoribbon Transistors with Ultrathin High-κ Dielectrics. *Nano Lett.* **10**, 1917-1921 (2010).

23. Liao, L. et al. Sub-100 nm Channel Length Graphene Transistors. *Nano Lett.* **10**, 3952-3956 (2010).

24. Cheng, R. et al. Few-layer molybdenum disulfide transistors and circuits for high-speed flexible electronics. *Nat. Commun.* **5**, 5143 (2014).

25. Wu, B. et al. Precise, Self-Limited Epitaxy of Ultrathin Organic Semiconductors and Heterojunctions Tailored by van der Waals Interactions. *Nano Lett.* **16**, 3754-3759 (2016).

26. Hersam, M.C. & Wang, Q.H. Room-temperature molecular-resolution characterization of self-assembled organic monolayers on epitaxial graphene. *Nat. Chem.* **1**, 206-211 (2009).

27. Zhao, Y., Wu, Q., Chen, Q. & Wang, J. Molecular Self-Assembly on Two-Dimensional Atomic Crystals: Insights from Molecular Dynamics Simulations. *J. Phys. Chem. Lett.* **6**, 4518-4524 (2015).

28. Martin, J.M., Vacher, B., Ponsonnet, L. & Dupuis, V. Chemical bond mapping of carbon by image-

spectrum EELS in the second derivative mode. *Ultramicroscopy* **65**, 229-238 (1996).

29. Wang, X., Xu, J., Wang, C., Du, J. & Xie, W. High-Performance Graphene Devices on $SiO_2$/Si Substrate Modified by Highly Ordered Self-Assembled Monolayers. *Adv. Mater.* **23**, 2464-2468 (2011).

30. Zou, X. et al. Interface Engineering for High-Performance Top-Gated $MoS_2$ Field-Effect Transistors. *Adv. Mater.* **26**, 6255-6261 (2014).

31. Wang, B. et al. High-κ Gate Dielectrics for Emerging Flexible and Stretchable Electronics. *Chem. Rev.* **118**, 5690-5754 (2018).

32. Cheng, R. et al. High-frequency self-aligned graphene transistors with transferred gate stacks. *Proc. Natl Acad. Sci. USA* **109**, 11588-11592 (2012).

33. Wu, Y. et al. High-frequency, scaled graphene transistors on diamond-like carbon. *Nature* **472**, 74-78 (2011).

34. Yu, Z. et al. Realization of Room-Temperature Phonon-Limited Carrier Transport in Monolayer $MoS_2$ by Dielectric and Carrier Screening. *Adv. Mater.* **28**, 547-552 (2016).

35. International Technology Roadmap for Semiconductors. www.itrs.net/2013-itrs.html (Semiconductor Industry Association, 2013).

36. Pu, J. et al. Highly Flexible and High-Performance Complementary Inverters of Large-Area Transition Metal Dichalcogenide Monolayers. *Adv. Mater*. **28**, 4111-4119 (2016).

37. Fang, N. & Nagashio, K. Band tail interface states and quantum capacitance in a monolayer molybdenum disulfide field-effect-transistor. *J. Phys. D* **51**, 65110 (2018).

38. Liu, H. & Ye, P.D. $MoS_2$ Dual-Gate MOSFET With Atomic-Layer-Deposited $Al_2O_3$ as Top-Gate Dielectric. *IEEE Electr. Dev. Lett.* **33**, 546-548 (2012).

39. Salvatore, G.A. et al. Fabrication and Transfer of Flexible Few-Layers $MoS_2$ Thin Film Transistors to Any Arbitrary Substrate. *ACS Nano* **7**, 8809-8815 (2013).

40. Ninomiya, N. et al. Fabrication of high-k/metal-gate $MoS_2$ field-effect transistor by device isolation process utilizing Ar- plasma etching. *J. Appl. Phys. Jpn* **54**, 046502 (2015).

41. Choi, K. et al. Trap density probing on top-gate MoS$_2$ nanosheet field-effect transistors by photo-excited charge collection spectroscopy. *Nanoscale* **7**, 5617 (2015).

42. Zhong, D. et al. Gigahertz integrated circuits based on carbon nanotube films. *Nat. Electron.* **1**, 40-45 (2018).

43. Auth, C. et al. A 22nm high performance and low-power CMOS technology featuring fully-depleted tri-gate transistors, self-aligned contacts and high density MIM capacitors. *Symp. VLSI Tech.* 131-132 (2012).

44. Wang, L. et al. One-dimensional electrical contact to a two-dimensional material. *Science* **342**, 614-617 (2013).

45. Takahashi, N. & Nagashio, K. Buffer layer engineering on graphene via various oxidation methods for atomic layer deposition. *Appl. Phys. Express* **9**, 125101 (2016).

46. Xia, Jilin, et al. Measurement of the quantum capacitance of graphene. *Nat. Nanotechnol.* **4**, 505-509 (2009).

47. McPherson, J., Kim, J., Shanware, A. & Mogul, H. Thermochemical description of dielectric breakdown in high dielectric constant materials. *Appl. Phys. Lett.* **82**, 2121-2123 (2003).

48. Jeong, S. et al. Thickness scaling of atomic-layer-deposited HfO$_2$ films and their application to wafer-scale graphene tunnelling transistors. *Sci. Rep.* **6** 20907 (2016).

49. Gusev, E.P. et al. Ultrathin high-κ gate stacks for advanced CMOS devices. *Inter. Electron Dev. Meet. Tech. Digest* 20.1.1-20.1.4 (2001).

50. Johannes Muller. et al. Ferroelectricity in simple binary ZrO$_2$ and HfO$_2$. *Nano Lett.* **12**, 4318-4223(2012).

**Acknowledgements.** This work is supported by National Natural Science Foundation of China 61734003, 61521001, 51861145202, 61861166001, 11874199, 21872100; National Key Basic Research Program of China 2015CB921600, 2015CB654901; Natural Science Foundation of Jiangsu Province under BK20170005; Singapore MOE Grant R143-000-A43-114; the Program A for Outstanding Ph.D. candidate of Nanjing University 201801A013; Postgraduate Research

& Practice Innovation Program of Jiangsu Province KYCX18_0045; Strategic Priority Research Program of Chinese Academy of Sciences XDB 30000000; a Grant-in-Aid for JSPS Research Fellows from the JSPS KAKENHI; Key Laboratory of Advanced Photonic and Electronic Materials, Collaborative Innovation Center of Solid-State Lighting and Energy-Saving Electronics, and the Fundamental Research Funds for the Central Universities, China.

**Author Contributions.** X.Wang conceived and supervised the project. W.L., J. Zhou, Z.Y., N.D., X.Wu., H.Z., D.H. Y.S. and X.D. contributed to sample preparation, characterization, device fabrication, measurements and data analysis. S.C. and P.W. performed TEM and data analysis. J.Zhang. and W.C. performed STM and data analysis. N.F. and K.N. performed $D_{it}$ analysis. T.L. and Z.W. performed CVD sample growth and transfer. Y.W. and T.C. contributed to RF transistor fabrication, measurements and data analysis. X.X. and H.M. performed molecular dynamics simulations. K.Y. and L.P. performed water contact angle measurement. W.L., Z.Y., P.W. and X.Wang wrote the manuscript with input from other authors. All authors contributed to discussions.

**Competing financial interests**

The authors declare no competing financial interests.

**Methods**

**Growth of ML PTCDA on 2D materials.** The growth of ML PTCDA was carried out in a home-built single-zone vacuum tube furnace. A crucible containing PTCDA powder (95%, Alfa Aesar) was placed in the center of the heating zone. Exfoliated or CVD 2D materials were placed downstream several centimeters away from the center. A turbo molecular pump was used to evacuate the quartz tube to ~$4 \times 10^{-5}$ Pa. Then, the source powder was heated to a target temperature of 250-280 °C (depending on the 2D materials) for ~20 minutes before cooling down naturally.

**ALD of HfO₂.** ALD was carried out at 150°C and base pressure of ~5 Pa using Tetrakis(dimethylamido) hafnium and H$_2$O as precursors. We used 30 sccm N$_2$ as carrier gas. The pulse/purge time for Hf and H$_2$O precursors were 250 ms/60 s and 100 ms/60 s, respectively.

**STM Characterization.** STM measurements were carried out at 77 K in an ultrahigh vacuum chamber (Unisoku LT-STM) with base pressure better than $1.0 \times 10^{-10}$ mbar. The STM imaging were performed using constant current mode with a commercial Pt-Ir tip. All the bias voltages were applied to the sample. For better conductivity, we used monolayer CVD graphene on Cu as the substrate, which was thoroughly degassed at 500 °C in the UHV chamber before in-situ PTCDA deposition. PTCDA are thermally evaporated from Knudsen cells at 280 °C onto the clean graphene/Cu substrate kept at 105 °C. Prior to deposition, the molecular source was purified by gradient vacuum sublimation.

**Calculation of binding energy.** Molecular mechanics method was applied to calculate the binding energies of PTCDA molecules on 2D materials via Forcite module in Materials Studio 7.0 package (Accelrys) by $E_{BE} = E_{PTCDA} + E_{sub} - E_{PTCDA@sub}$. Here $E_{PTCDA@sub}$, $E_{PTCDA}$ and $E_{sub}$ are total energy of PTCDA molecule adsorbed on the substrate, isolated PTCDA and substrate, respectively. In this case, universal force field (UFF) was used.

**Cross-sectional TEM sample fabrication and STEM characterization.** TEM samples of Au/HfO$_2$/ML PTCDA/2D materials structure were fabricated using FEI Helios 600i dual-beam focused ion beam (FIB) system. Pt protection layer was first deposited on the top surface of the devices, followed by etching surrounding area to form the sample lamella using 30 kV accelerating voltage and 21 nA gallium ion beam. The lamella was then lift out from silicon substrate and directly transferred to a TEM half grid inside the FIB chamber. We further thinned the observation area down to less than 100 nm with tens to hundreds pA gallium beam current. Finally, 5 kV accelerating voltage with 0.12 nA and 2 kV accelerating voltage with a smaller current of 68 pA was subsequently used for fine polishing to remove the damage layers present on both sides of the prepared TEM samples.

STEM characterization was acquired on a FEI Titan G2 60-300 aberration corrected S/TEM microscope with the accelerating voltage of 300kV. Bright-field (BF) imaging mode was used for the device observation. Unlike Z-contrast image, a BF image shows phase contrast, so it is much more applicable to distinguish light atoms from heavy ones. Gatan dual-EELS system was used for STEM-EELS data acquisition. For EELS mapping, the acquisition time per pixel was 0.1 s to reduce beam damage. The total acquisition energy range was 60 to 572 eV, including carbon *K*-edge, sulfur *L*-edge and oxygen *K*-edge. By integrating the signal of corresponding energy loss edges from STEM-EELS images, the distribution of those elements can be mapped.

**CVD MoS$_2$ growth and transfer.** Centimeter-size MoS$_2$ films were grown by the way of low-pressure CVD on sapphire substrate using MoO$_3$ powder and S powder as sources, under 950 °C. To transfer CVD MoS$_2$ film off sapphire using ML PTCDA/HfO$_2$ stack, we first grew ML PTCDA and HfO$_2$ on freshly grown CVD samples using preceding method. PMMA was spin coated at a speed of 2000 R.P.M. for 1 min and then cured at 70 °C for 10 min, followed by lamination of thermal release tape. Next, the sample was immersed in deionized water and the whole stack was spontaneously delaminated from sapphire and transferred onto a target substrate. Finally, the thermal release tape was removed by baking at 150 °C, and PMMA was removed by soaking in acetone.

**Topgate transistor fabrication.** All the topgate devices shared similar fabrication procedures. First, source/drain electrodes were patterned. Then, ML PTCDA and HfO$_2$ were deposited onto the channel area using preceding method. Finally, gate electrode was align patterned between source and drain. For exfoliated graphene/TMD transistors, 80 nm Au was used as source/drain/gate electrodes. For graphene RF transistors, 5 nm Ti/70 nm Pd was used as source/drain/gate electrodes to enhance adhesion and reduce parasitic resistance. For CVD MoS$_2$ transistors, 10 nm Ti/50 nm Au (5 nm Ti/15 nm Au) was used as source/drain (gate) electrodes.

**Short-channel MoS$_2$ transistor fabrication.** Few-layer graphene was mechanically exfoliated onto 275 nm SiO$_2$/Si substrate. Then, Au was contacted to the graphene as gate electrode. ML PTCDA and 6 nm HfO$_2$ was deposited onto graphene as gate dielectric using preceding methods. MoS$_2$ was align transferred onto HfO$_2$/ML PTCDA/graphene stack using pole-propylene carbonate film. Source and drain electrodes were patterned by EBL. We deposited 1 nm Ti/10 nm Pd metal electrodes to minimize the fringing field and SCE.

**Electrical measurements.** Electrical measurements of graphene and TMD devices were carried out by an Agilent B1500 semiconductor parameter analyzer in a close-cycle cryogenic probe station with base pressure ~10$^{-6}$ Torr. *C-f* measurements were performed by Agilent

E4980 Precision LCR Meter. The on-chip microwave measurements of graphene RF transistors were carried out by using Agilent N5247A network analyzer under ambient environment.

# Supplementary Information

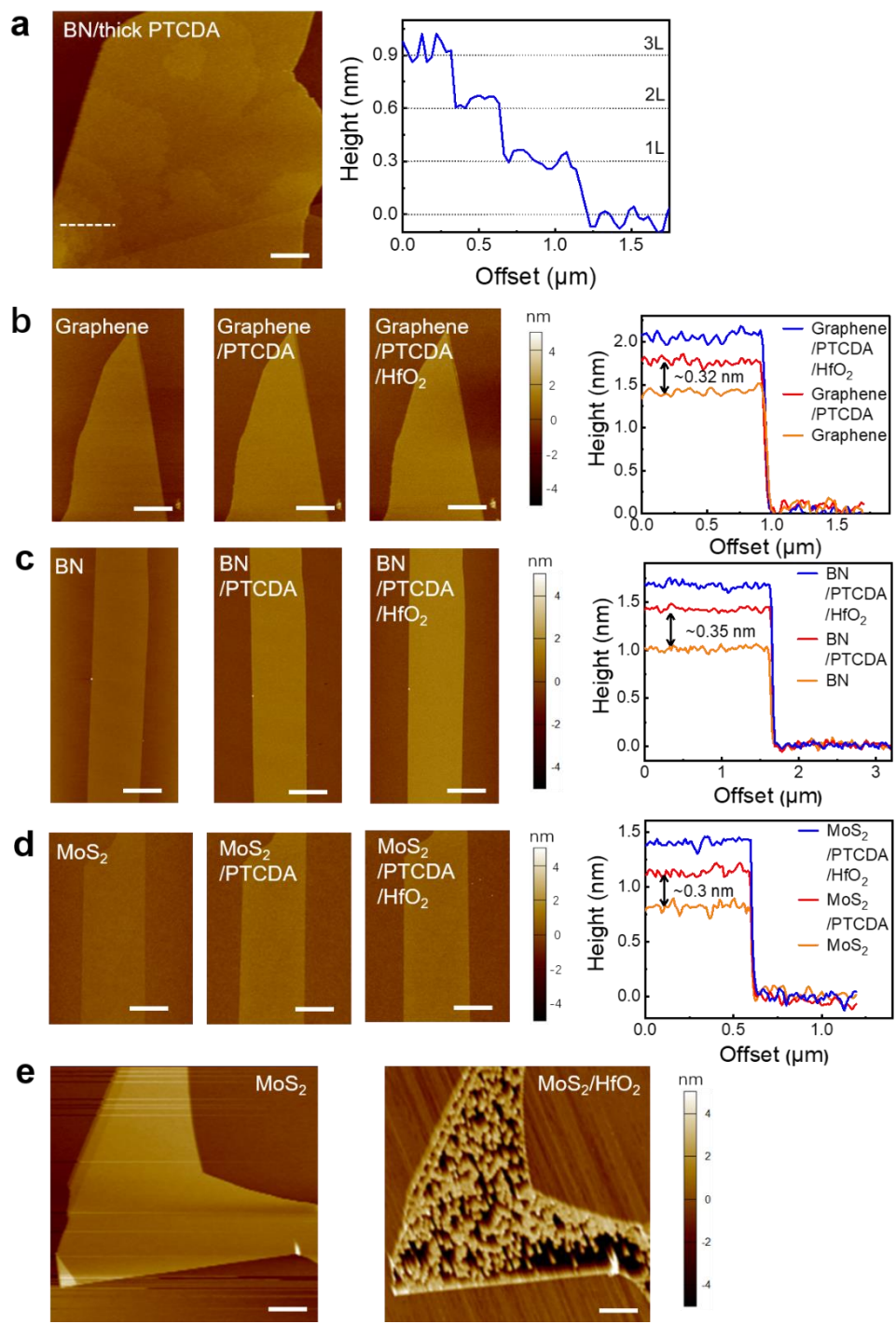

**Supplementary Figure 1. AFM characterization. a.** Left, AFM image of a multi-layer PTCDA sample on BN showing layered morphology (left). Scale bar: 4 μm. Right, height profile measured from the AFM image. The thickness of each layer is ~0.3 nm. **b-d.** AFM images of exfoliated 2D materials (graphene, BN, and MoS$_2$, respectively, left row), after ML PTCDA deposition (middle row) and after 2 nm HfO$_2$ deposition (right row). Scale bars: 3 μm. The height profiles after each step are shown on the right. **e.** AFM images of an exfoliated MoS$_2$ (left) and after 10 nm HfO$_2$ deposition without PTCDA (right). Scale bars: 3 μm.

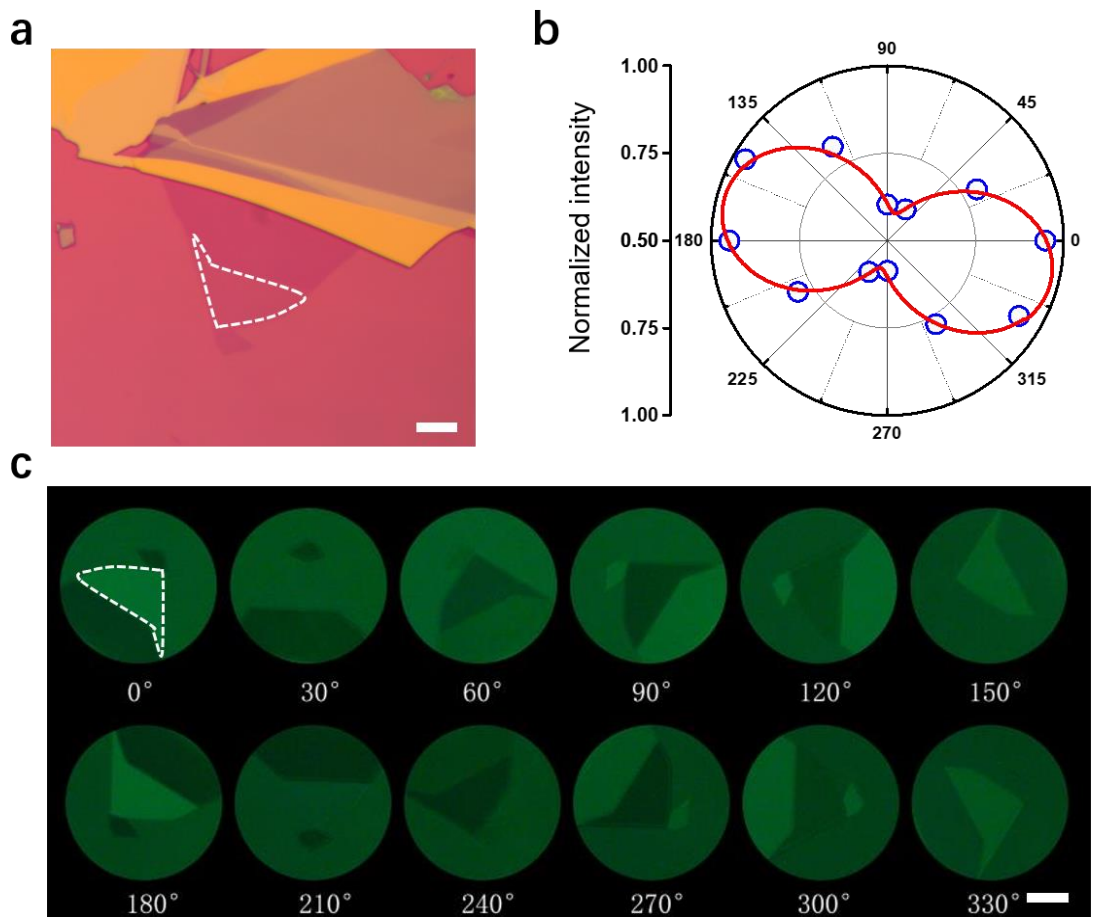

**Supplementary Figure 2. Cross-polarized optical microscopy. a.** Optical microscopic image of a ML PTCDA/graphene sample. The highlighted area is monolayer graphene. Scale bar: 5 $\mu$m. **b.** Polar plots of the normalized intensity under cross-polarized optical microscope measured from **c**. **c.** Cross-polarized microscopic images of the sample at different rotation angles under 532 nm illumination. As we rotate the sample, the intensity is uniformly modulated with 180 ° period, which is consistent with the symmetry of PTCDA packing and confirms the single-crystalline nature of the ML PTCDA. Scale bar: 5 $\mu$m.

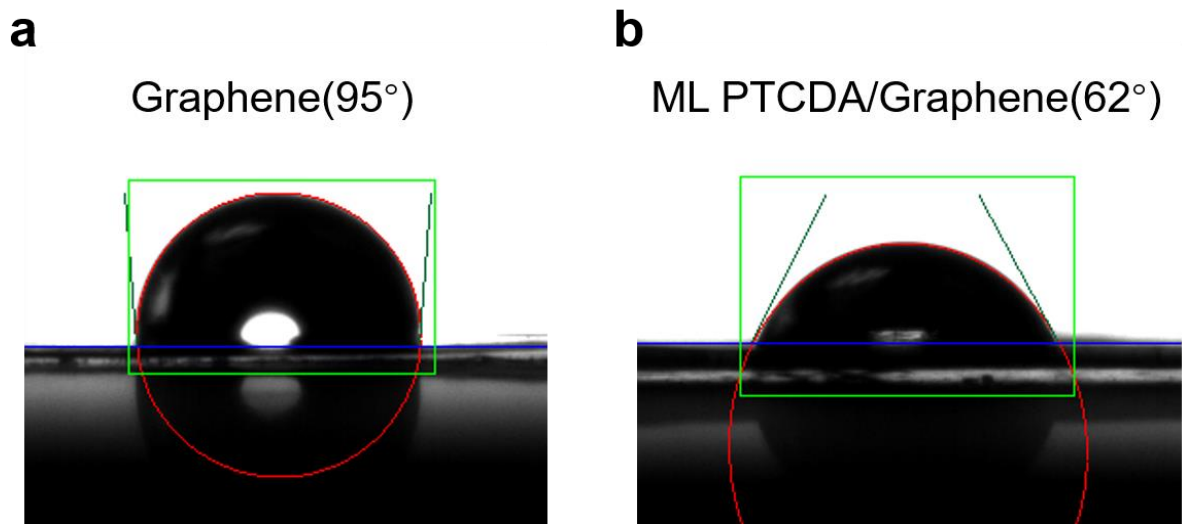

**Supplementary Figure 3. Water contact angle measurement.** The water contact angle of CVD graphene on Cu foil before (**a**) and after (**b**) deposition of ML PTCDA. The contact angle decreased from 95° to 62°.

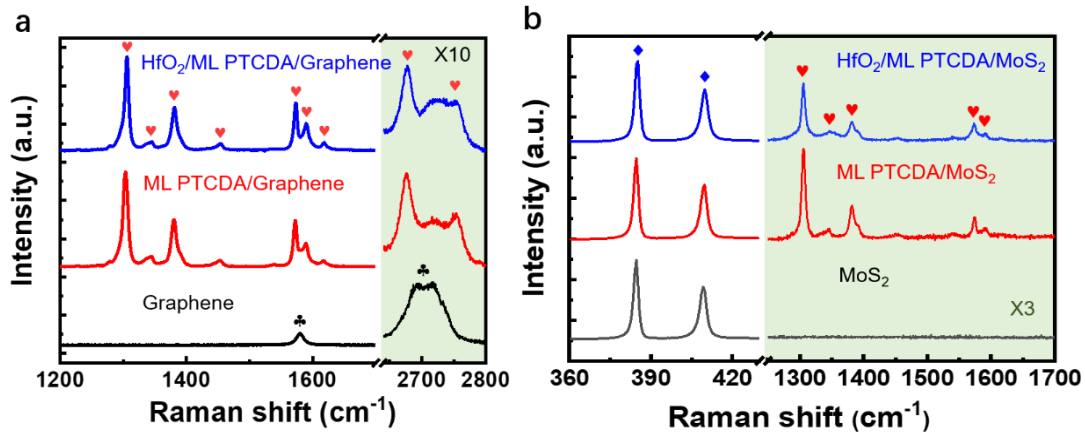

**Supplementary Figure 4. Raman characterization. a.** (From bottom to top) The Raman spectrum of an exfoliated few-layer graphene, after ML PTCDA and after ALD $HfO_2$. **b.** (From bottom to top) The Raman spectrum of an exfoliated few-layer $MoS_2$, after ML PTCDA and after ALD $HfO_2$. Red Heart symbol represent the characteristic Raman peak of PTCDA (1305 cm$^{-1}$, 1345 cm$^{-1}$, 1381 cm$^{-1}$, 1573 cm$^{-1}$, 1590 cm$^{-1}$, 1618 cm$^{-1}$, 2686 cm$^{-1}$, 2757 cm$^{-1}$). Black Club symbol represent the characteristic Raman peak of graphene (1580 cm$^{-1}$, 2710 cm$^{-1}$). Blue Diamond symbol represent the characteristic Raman peak of $MoS_2$ (385 cm$^{-1}$, 409 cm$^{-1}$). We can see that in both cases, the PTCDA layer is still well preserved during ALD process.

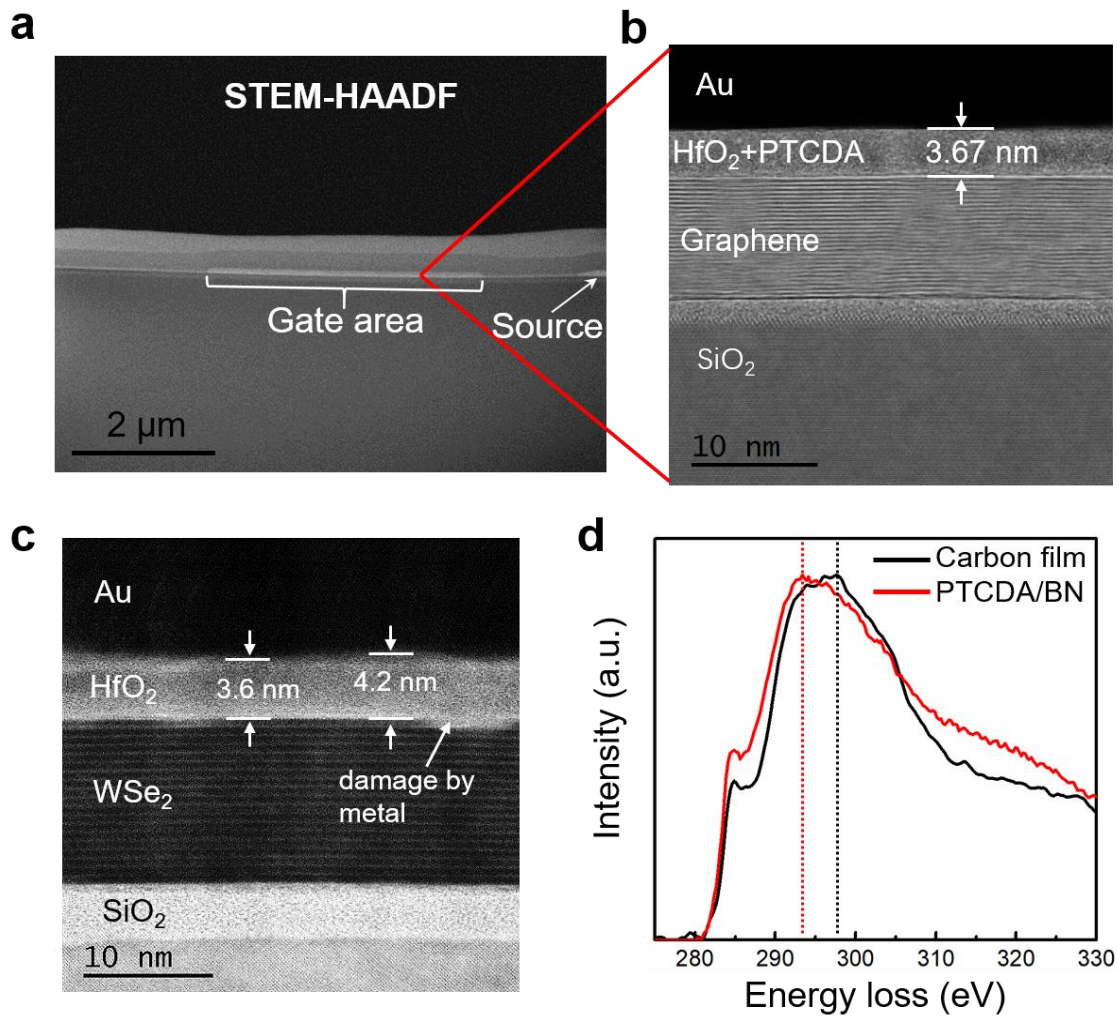

**Supplementary Figure 5. Cross-sectional STEM characterization. a-b.** Cross-sectional STEM high angle annular dark field (STEM-HAADF) at low magnification (a) and high-resolution STEM-BF (**b**) images of an Au/HfO$_2$/ML PTCDA/multi-layer graphene stack. The topgate and source metal electrodes are marked in (**a**). The thin white horizontal line across the entire image in (**a**) is the uniform deposition of ML PTCDA/HfO$_2$. **c.** Cross-sectional STEM-BF image of an Au/HfO$_2$/1 nm Al/WSe$_2$ stack, showing large thickness variation of 0.6 nm in HfO$_2$. In addition, damage of top WSe$_2$ layer by Al metal is evident. **d.** EELS spectra of the PTCDA grown on BN (Red) and amorphous carbon (Black) on carbon film of a TEM grid, respectively, which is consistent with the result shown in Fig. 2e.

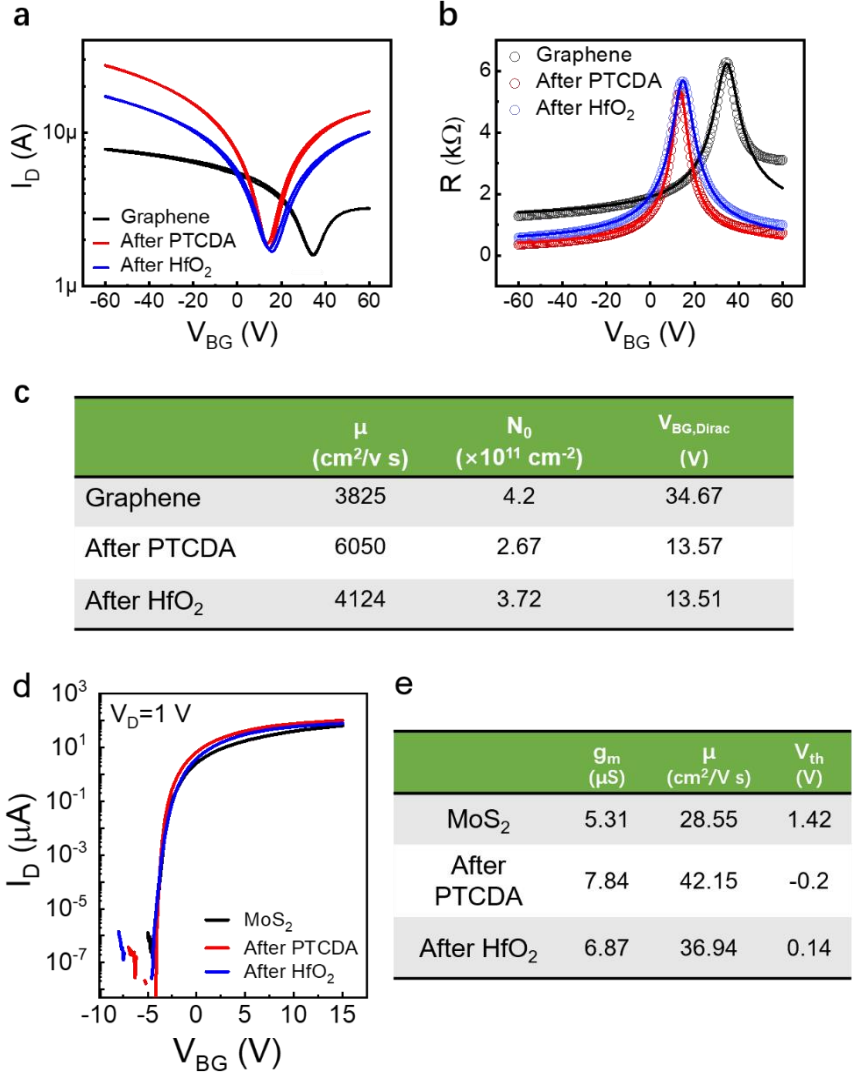

**Supplementary Figure 6. Step-by-step electrical measurements of topgate FET. a.** Double-sweep transfer curves of a backgate graphene FET on 275 nm SiO$_2$/Si substrate as fabricated, after growing ML PTCDA and after 4 nm HfO$_2$ ALD. $V_D$=10 mV. **b.** Experimental data (symbols) and numerical fitting (lines) of $R$ vs. $V_{BG}$ from (**a**) to derive the mobility and residual carrier density. The fitting procedure[1] is as follow: the carrier density $n_{TG}$ in the channel can be described by the gate capacitance and the quantum capacitance. The expression is $V_{TG} - V_{TG,Dirac} = \frac{e}{C_{ox}} n_{TG} + \frac{\hbar v_F \sqrt{\pi n_{TG}}}{e}$, where $C_{ox}$ is the gate capacitance and $v_F$ is the Fermi velocity. The total resistance consists of the contact resistance and channel resistance and can be described by $R_{total} = R_c + R_{ch} = R_c + \frac{L/W}{\sqrt{n_0^2 + n_{TG}^2} e\mu}$. Where $L/W$, $\mu$ and $n_0$ are Length/Width ratio, mobility and residual carrier density induce by charged impurities, respectively. **c.** The derived mobility and residual carrier density after each step. The mobility and residual carrier density are improved and the Dirac point is closer to zero after the fabrication of gate stack, which show that PTCDA does not introduce defects, doping or interface states in graphene. **d.** Transfer curves of a backgate $MoS_2$ FET on 30 nm $Al_2O_3$/Si substrate as fabricated, after growing ML PTCDA and after 4 nm $HfO_2$ ALD. $V_D$=1 V. **e.** The derived transconductance, mobility and threshold voltage after each step.

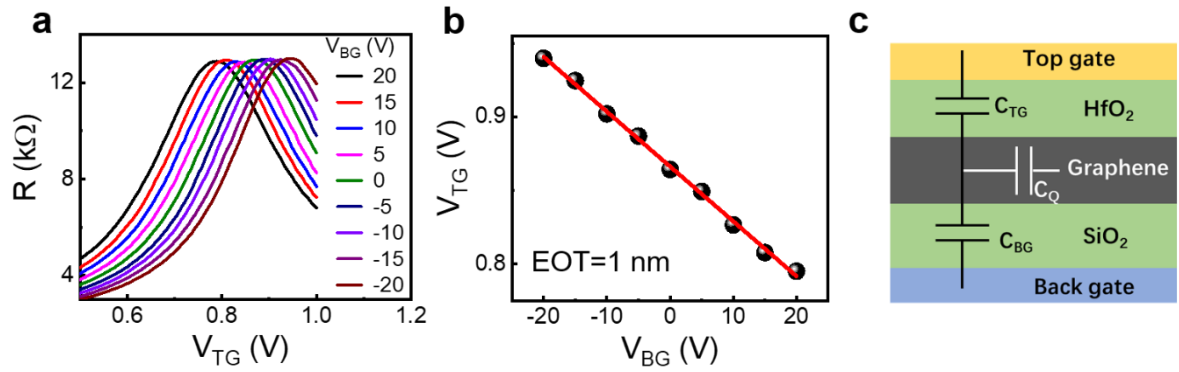

**Supplementary Figure 7. Double-gate measurement to extract $C_{TG}$. a.** Transfer curves of a topgate graphene FET under different $V_{BG}$ from -20 to 20 V, $V_D$=10 mV. **b.** Dirac point voltage as a function of $V_{BG}$. Solid line shows linear fitting of the experiment data. The slope is -0.0036, corresponding to EOT=1 nm. **c.** A schematic view of the dual-gate model[2]. $C_{TG}$ is the total capacitance of HfO$_2$ layer/PTCDA layer, $C_Q$ is the quantum capacitance of graphene and $C_{BG}$ is capacitance of 275 nm SiO$_2$.

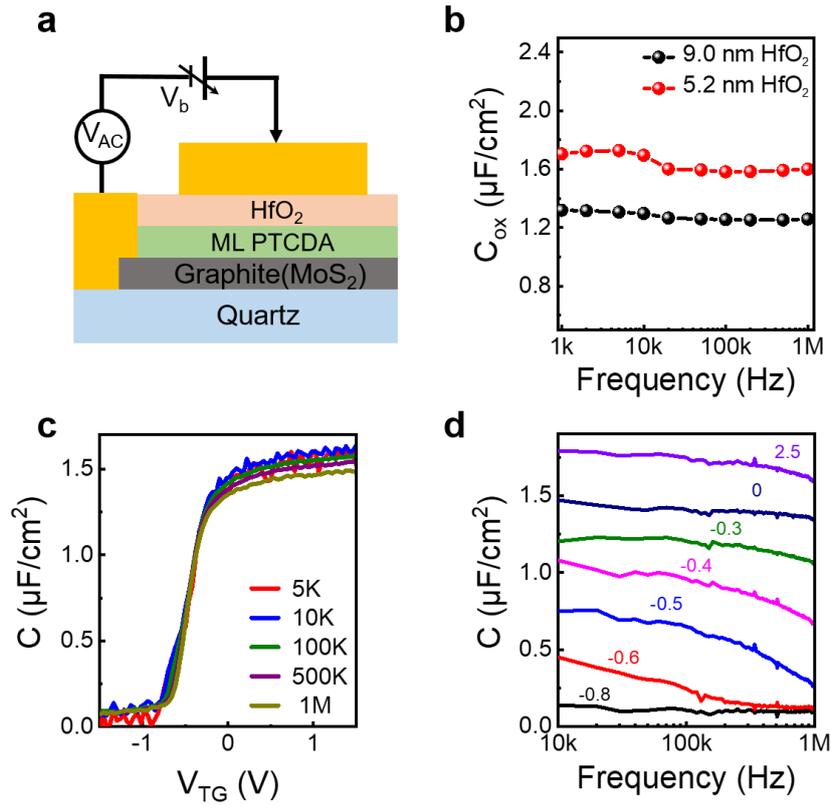

**Supplementary Figure 8. Capacitance-frequency (*C-f*) measurement and capacitance-voltage (*C-V*) measurement to extract $C_{TG}$. a.** Schematic of the experimental capacitor structure. **b.** Measured capacitance of HfO$_2$/PTCDA stack with two different HfO$_2$ thickness in the frequency range of 1 kHz – 1 MHz. **c.** *C-V* characteristics under different frequency with 6 nm HfO$_2$/MoS$_2$ structure on quartz substrate. **d.** *C-f* characteristics under different gate voltage $C_{TG}$ (V) with the same device in **c**.

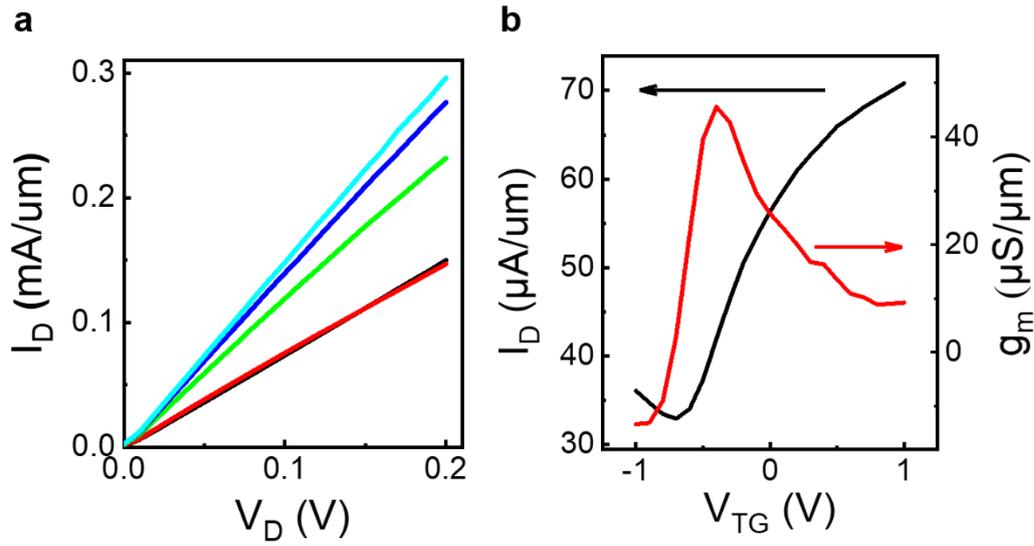

**Supplementary Figure 9. DC performance of the RF graphene transistor shown in Fig. 3 of main text. a.** Low bias $I_D$-$V_D$ output characteristics at various gate voltages ($V_{TG}$ from -1 to 1 V, step is 0.5 V). **b.** $I_D$-$V_{TG}$ transfer characteristics and transconductance at $V_D$=50 mV.

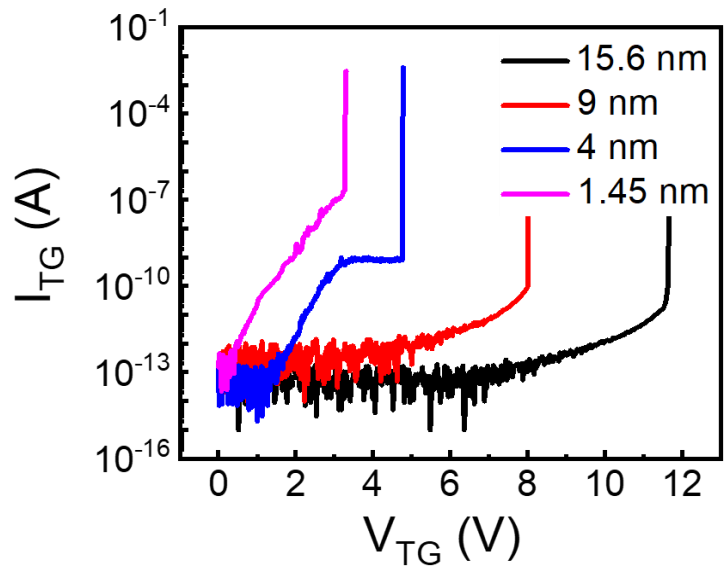

**Supplementary Figure 10. Breakdown characteristics of ML PTCDA/HfO₂.** Gate leakage current $I_{TG}$ as a function of $V_{TG}$ measured on topgate graphene FETs with different $t_{ox}$ of 1.45 nm, 4 nm, 9 nm and 15.6 nm.

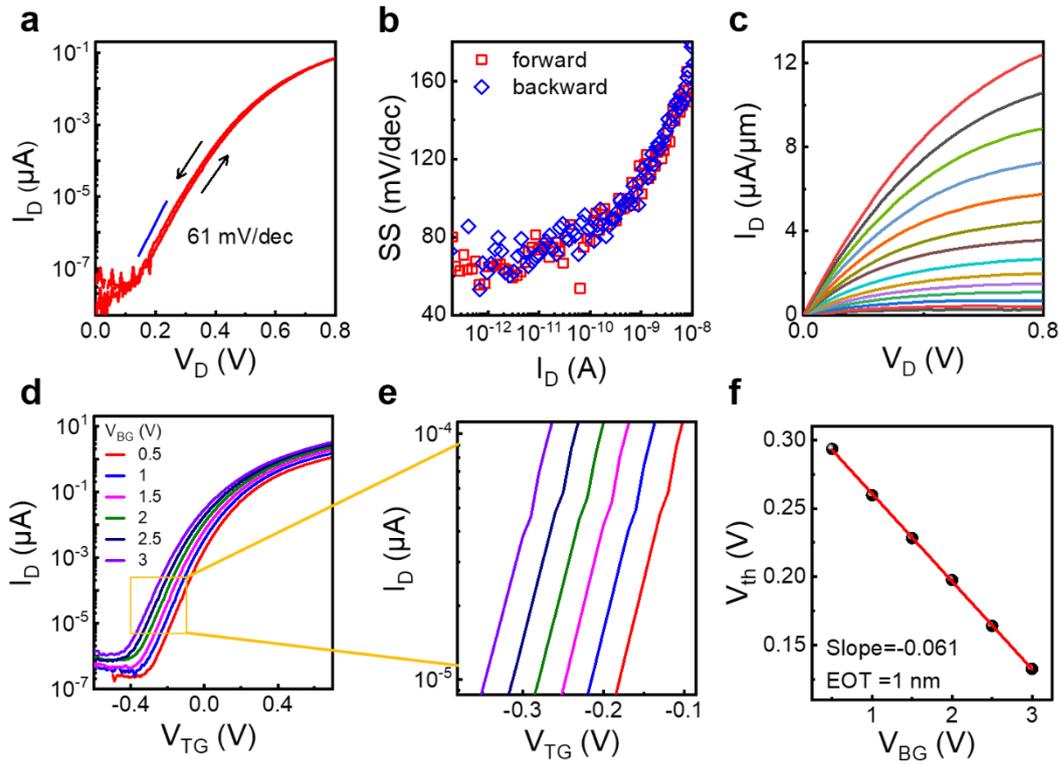

**Supplementary Figure 11. Topgate MoS$_2$ EFT with 1 nm EOT ($t_{ox}$=1.45 nm, monolayer MoS$_2$). a.** Double-sweep transfer characteristics measured under $V_D$=0.1 V and $V_{BG}$=-0.5 V, showing SS=61 mV/dec and no hysteresis. **b.** SS versus $I_D$ characteristics of the device in **a**, showing constant SS across over 3 orders of $I_D$ with minimum SS of ~60 mV/dec for both forward and reverse sweeping. **c.** Output characteristics of the same device under $V_{BG}$=15 V. From top to bottom, $V_{TG}$ from 0.8 V to -0.5 V with 0.1 V step. **d.** Transfer characteristics under different $V_{BG}$ from 0.5 V to 3 V with 0.5 V step. **e.** Magnification of the transfer curves from **d**. **f.** The threshold voltage $V_{th}$ as the function of $V_{BG}$. From the linear fitting, EOT=1 nm is obtained.

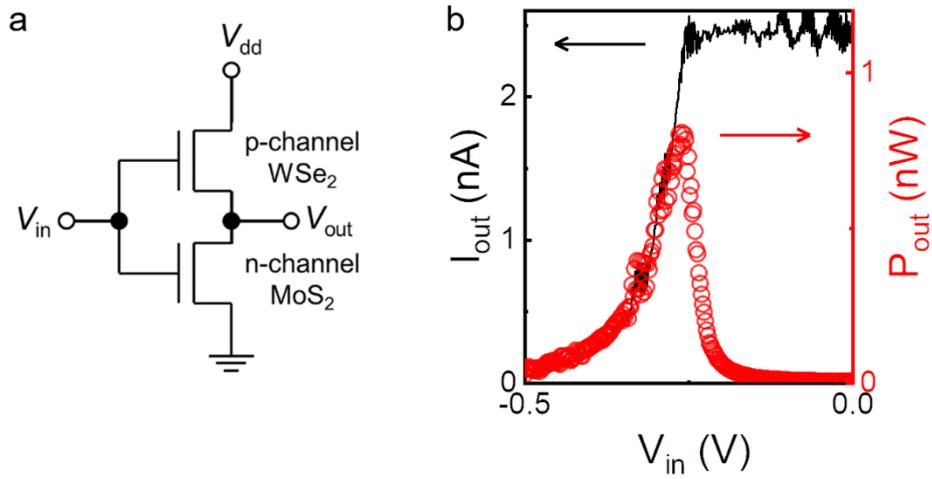

**Supplementary Figure 12. Ultra-low power CMOS inverter. a.** The equivalent circuit diagram. WSe$_2$ is used in a p-channel transistor and MoS$_2$ is used in a n-channel transistor, which are then connected in series to construct a CMOS inverter. **b.** Measured operating current $I_{dd}$ and Power $P_{out}$ as function of $V_{in}$ at $V_{dd}$=0.5 V.

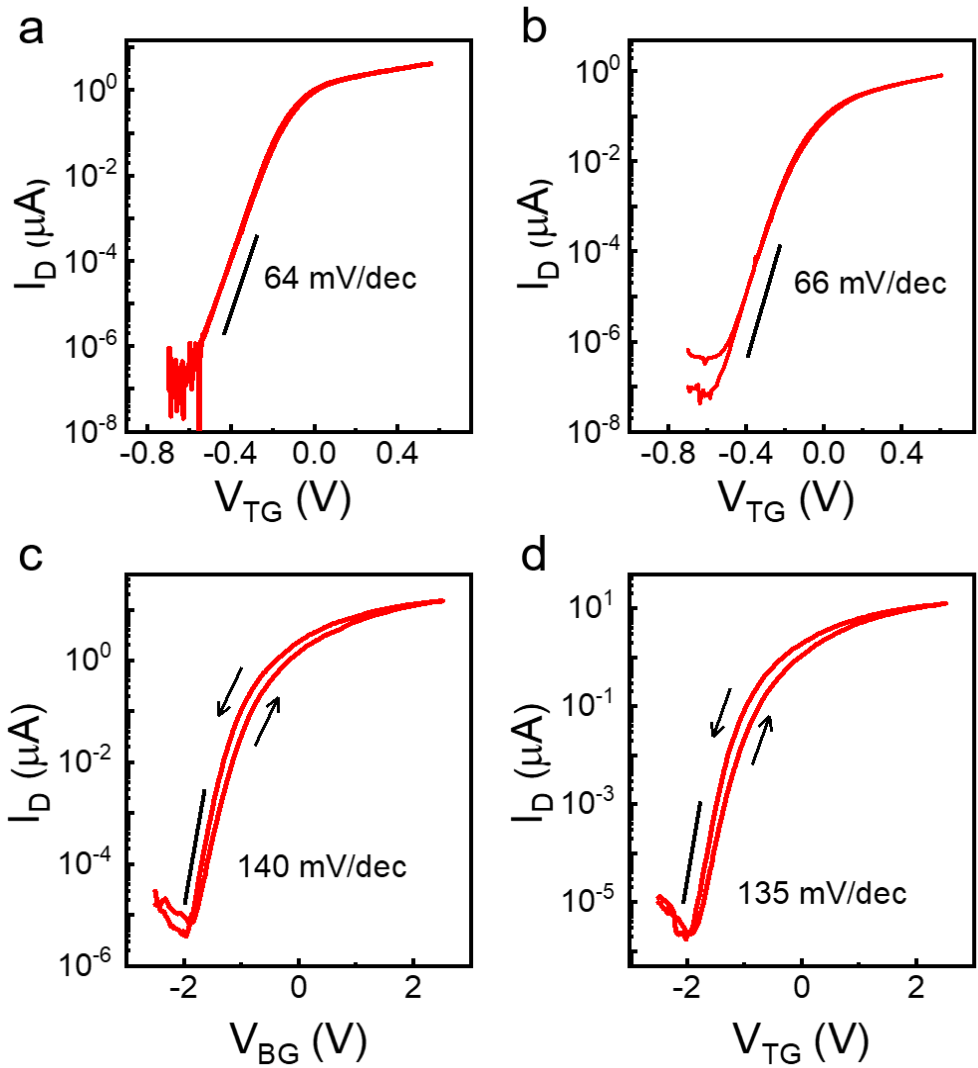

**Supplementary Figure 13. Hysteresis of topgate MoS$_2$ FET. a-b**. Double-sweep transfer characteristics of two typical FETs based on exfoliated MoS$_2$ samples measured under $V_D$=0.5 V with negligible hysteresis (<10 mV). **c-d.** Double-sweep transfer characteristics of two typical FETs based on CVD MoS$_2$ samples measured under $V_D$=0.1 V, showing slight hysteresis (~100 mV) which may be induced by PMMA residual during transfer processes.

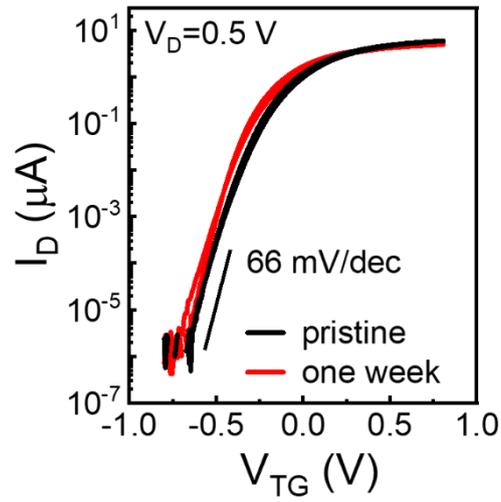

**Supplementary Figure 14. Long-term stability.** Transfer curves of topgate MoS$_2$ transistor with 4 nm HfO$_2$ as gate dielectric. The black line and red line show the results from the same device after fabrication and after one-week storage, respectively.

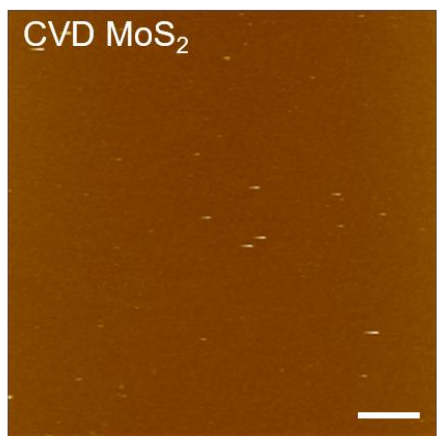 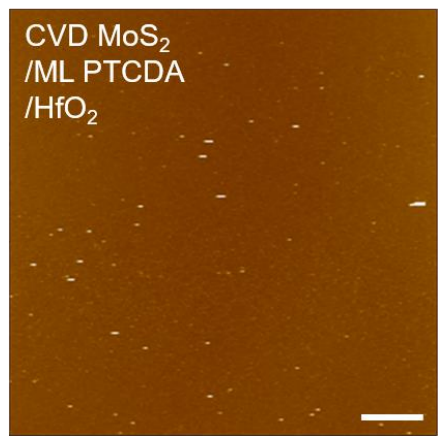

**Supplementary Figure 15. AFM characterization of CVD MoS$_2$ film.** AFM images of transferred CVD MoS$_2$ film before (**a**) and after (**b**) depositing ML PTCDA/6 nm HfO$_2$. Roughness are 0.2 nm and 0.3 nm respectively. Scale bars: 3 μm.

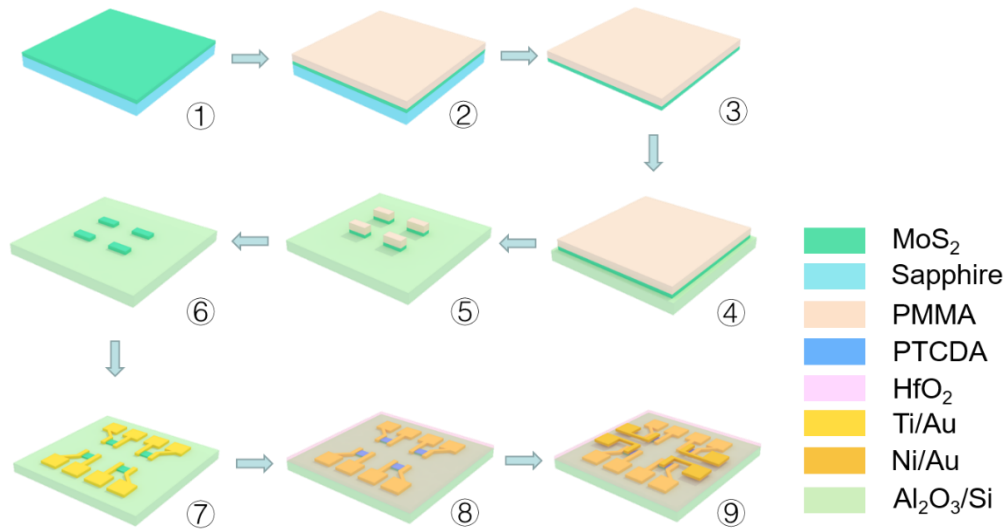

**Supplementary Figure 16. Fabrication process of topgate FET array on CVD MoS$_2$**

① CVD growth of monolayer MoS$_2$ film on sapphire.

② Spin coat PMMA as a transfer support layer.

③ Use 30% NaOH solution to etch the sapphire substrate until PMMA/MoS$_2$ layer separating from sapphire substrate.

④ After several times of deionized water cleaning, the PMMA/MoS$_2$ layer was transferred to 30 nm Al$_2$O$_3$/Si.

⑤ Pattern the channel regions on MoS$_2$ film by electron beam lithography (EBL) and CF$_4$ plasma etching.

⑥ Remove PMMA using acetone.

⑦ Pattern source/drain electrodes by EBL and deposit 10 nm Ti/50 nm Au by e-beam evaporation(EBE), then lift off.

⑧ Deposit ML PTCDA and 6 nm HfO$_2$.

⑨ Pattern topgate electrodes by EBL, deposition of 5 nm Ti/15 nm Au by EBE and lift off.

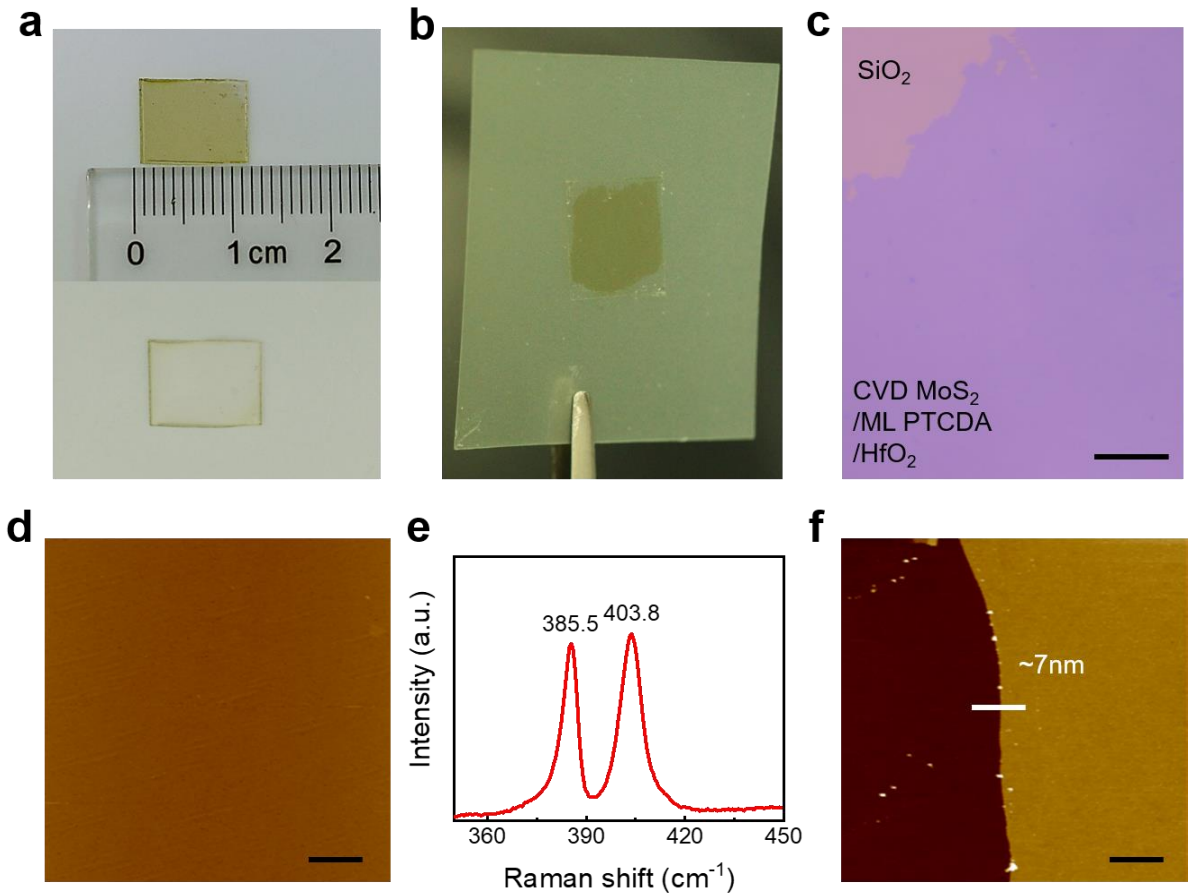

**Supplementary Figure 17. Transfer of CVD MoS$_2$ using ML PTCDA/HfO$_2$ stack as support layer. a.** Top image shows HfO$_2$/ML PTCDA deposited on centimeter MoS$_2$ film which is grown on sapphire. The bottom image shows sapphire substrate after transfer. Changes in color and transparency indicate that the CVD MoS$_2$ film has been successfully transferred. **b.** Photograph of CVD MoS$_2$ film/ML PTCDA/HfO$_2$ stack attached to thermal release tape during the transfer process. **c.** Optical microscope image of CVD MoS$_2$ film/ML PTCDA/HfO$_2$ stack transferred onto SiO$_2$/Si substrate. Scale bar is 200 μm. **d.** AFM image of the transferred film on SiO$_2$. Scale bar is 4 μm. **e.** Raman spectrum of the transferred film showing characteristic peak of monolayer MoS$_2$ film at 385.5 cm$^{-1}$ and 403.8 cm$^{-1}$. **f.** AFM image on the edge of the transferred film, showing ultra-clean surface and small roughness. We deposited 6nm HfO$_2$, so the total height of the MoS$_2$/ML PTCDA/HfO$_2$ stack is about ~7nm. Scale bar is 2 μm.

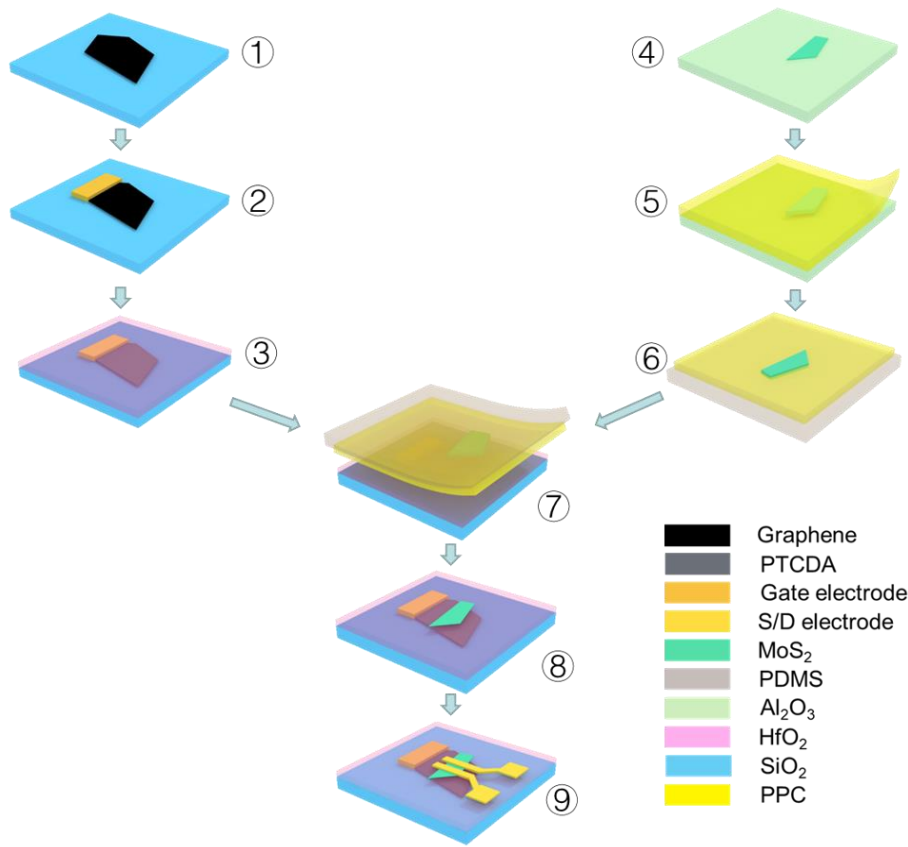

**Supplementary Figure 18. Fabrication process of short-channel MoS$_2$ FET**

① Exfoliate few-layer graphene on 275 nm SiO$_2$/Si substrate.

② Deposit Au electrode to cover a small portion of graphene as gate electrode.

③ Deposit ML PTCDA/HfO$_2$ on graphene as gate dielectric.

④ Exfoliate bi-layer MoS$_2$ flake on 30 nm Al$_2$O$_3$/Si substrate.

⑤ Spin coat pole-propylene carbonate (PPC) (Sigma-Aldrich, CAS 25511-85-7) layer on MoS$_2$.

⑥ The PPC/MoS$_2$ flake layer was peeled from the Al$_2$O$_3$/Si substrate and placed onto a transparent elastomer stamp (poly dimethyl siloxane, PDMS) for transfer.

⑦ The MoS$_2$ was transferred on the top of gate stack using a micro-manipulator under microscope.

⑧ Remove PPC with acetone.

⑨ Pattern source/drain electrodes by EBL and deposit 1 nm Ti/10 nm Pd by EBE, then lift off.

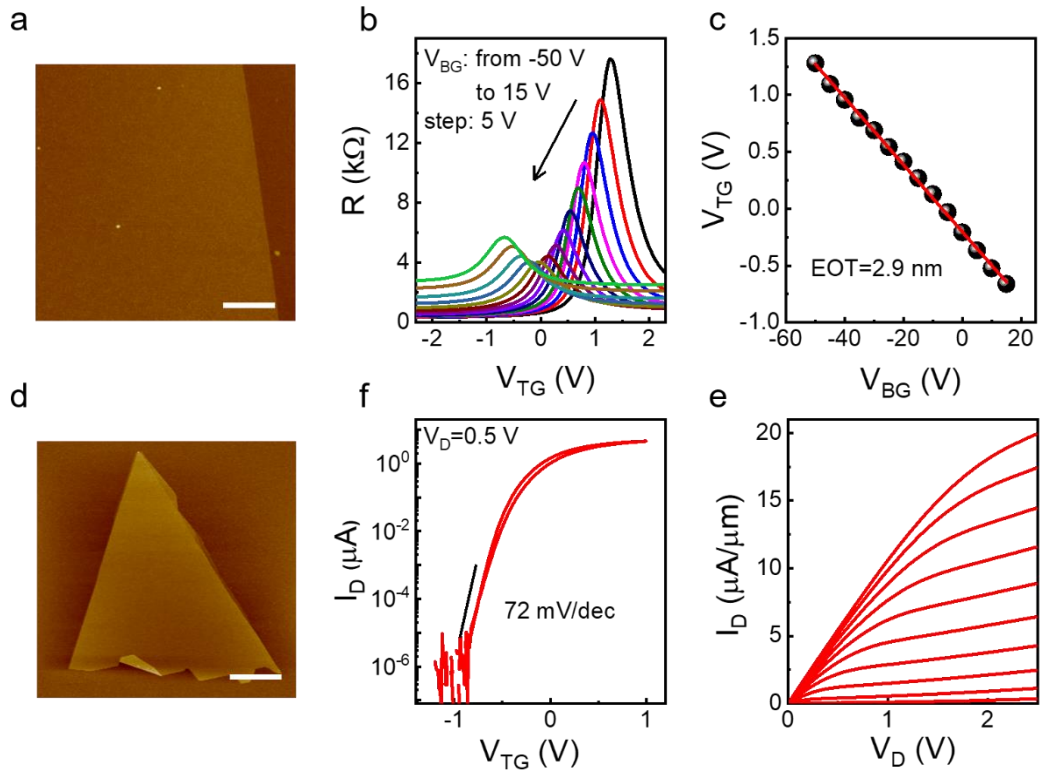

**Supplementary Figure 19. Topgate FETs with Al$_2$O$_3$ dielectric. a.** AFM image of 4 nm Al$_2$O$_3$/ML PTCDA/graphene structure on 100 nm SiO$_2$/Si substrate with roughness of ~150 pm. Scale bar is 2 μm. **b.** Transfer curves of a topgate bi-layer graphene FET ($t_{ox}$=4 nm) under different $V_{BG}$ from -50 to 15 V, $V_D$=10 mV. **c.** Dirac point as a function of $V_{BG}$. Solid line shows linear fitting of the experiment data with slope of -0.029, corresponding to EOT=2.9 nm. **d.** AFM image of 4 nm Al$_2$O$_3$/ML PTCDA/MoS$_2$ structure on 30 nm Al$_2$O$_3$/Si substrate with roughness of ~150 pm. Scale bar is 3 μm. **e.** Double-sweep transfer characteristics of MoS$_2$ FET ($t_{ox}$=4.5 nm) measured under $V_D$=0.1 V, showing SS=72 mV/dec and negligible hysteresis. **f.** Output characteristics of the same device under $V_{BG}$=15 V. From top to bottom, $V_{TG}$ from 1.4 V to -0.6 V with 0.2 V step.

**Supplementary Table 1. Comparison of ALD oxide properties on graphene**

| Dielectric layer | Modified method | Thickness of seed layer (nm) | Roughness (nm) | $t_{ox}$ (nm) | Gate capacitance (µF/cm$^2$) | EOT (nm) | Ref. |
|---|---|---|---|---|---|---|---|
| $HfO_2$ | NFC polymer | 10 | | 20 | 0.18 | 19 | 3 |
| $Al_2O_3$ | $O_2$ plasma | 1 | 0.37 | 5 | 0.78 | 4.42 | 4 |
| $HfO_2$ | PTCDA | 0.3-0.6 | 0.8-1.1 | 10 | 0.89 | 3.87 | 5 |
| $Al_2O_3$ | Ozone | | | 4.5 | 0.92 | 3.74 | 6 |
| $Y_2O_3$ | Y seed layer | 1 | 0.3 | 6 | 1.16 | 2.97 | 7 |
| $HfO_2$ | Hf seed layer | 3 | 0.88 | 5 | 2.5 | 1.38 | 8 |
| $HfO_2$ | Electron beam irradiation | | 0.4 | 4 | 2.63 | 1.3 | 9 |
| **$HfO_2$** | **ML PTCDA** | **0.3** | **0.13** | **1.5** | **3.45** | **1** | **this work** |

**Supplementary Table 2. Comparison of ALD oxide properties on TMDs**

| Gate methods | Dielectric layer | EOT (nm) | Roughness (nm) | Channel length(nm) | SS (mV/dec) | Ref. |
|---|---|---|---|---|---|---|
| Directly ALD | 30 nm $HfO_2$ | | | long channel | 75 | 10 |
| Ozone pretreatment | 6 nm $HfO_2$ | 3.55 | | | 75 | 11 |
| Remote oxygen plasma pretreatment | 6.6 nm $Al_2O_3$ | | 0.58 | | 101 | 12 |
| Plasma pretreatment | 6.1 nm $Al_2O_3$ | 3.55 | | | 85 | 13 |
| Deposit Y, and oxidize as seed layer | 3-7 nm $Y_2O_3$ + 9 nm $HfO_2$ | 4.42 | | | 65 | 14 |
| PEALD | 4 nm $HfO_2$ | | 0.3 | | ~ 130 | 15 |
| AlN as seed layer | 1 nm AlN+ 5 nm $Al_2O_3$ | | 0.63 | | ~ 120 | 16 |
| **ML PTCDA as seed layer** | **1.5 nm $HfO_2$** | **1** | **0.13** | | **60** | **this work** |
| 1L TiOPc as seed layer ($WSe_2$) | 5.3 nm $Al_2O_3$ | 3 | 0.15 | | 80 | 17 |
| Directly ALD $ZrO_2$ ($WSe_2$) | 17.5 nm $ZrO_2$ | 5.46 | | | 60 | 18 |
| **ML PTCDA as seed layer ($WSe_2$)** | **2.8 nm $HfO_2$** | **1.3** | **0.13** | | **67** | **this work** |
| Transfer $Al_2O_3$/Al nanowires | 5 nm $Al_2O_3$ | 2.5 | | 30 | 80 | 19 |
| Directly ALD | 10 nm HfO2 | | | 15 | 205 | 20 |
| Deposit Ti, and oxidize as seed layer | 4 nm $HfO_2$ /1.2 nm $TiO_2$ | | 0.4 | 14 | 86.5 | 21 |
| Directly ALD on Pt/Ti gate electrode | 7.5 nm $HfO_2$ | 2.5 | | 10 | 200 | 22 |
| Transfer $Al_2O_3$/Al nanowires | 5 nm $Al_2O_3$ | 2.5 | | 10 | 250 | 19 |
| Transfer h-BN as gate dielectric layer | 4 nm BN | | | 9 | 93 | 23 |
| Transfer h-BN as gate dielectric layer | 2.5 nm BN | | | 4 | 208 | 23 |
| Directly ALD on SWCNT gate electrode | 5.8 nm $ZrO_2$ | | | 1 | 65 | 24 |
| **ML PTCDA as seed layer** | **6 nm $HfO_2$** | **2** | **0.13** | **20** | **73** | **this work** |

Note: the channel material is $MoS_2$ unless otherwise stated.

**References:**

1. Kim, S. et al. Realization of a high mobility dual-gated graphene field-effect transistor with $Al_2O_3$ dielectric. *Appl. Phys. Lett.* **94**, 62107 (2009).

2. Xu, H., et al., Quantum capacitance limited vertical scaling of graphene field-effect transistor. *ACS Nano* **5**, 2340-2347 (2011).

3. Farmer, D.B. et al. Utilization of a Buffered Dielectric to Achieve High Field-Effect Carrier Mobility in Graphene Transistors. *Nano Lett.* **9**, 4474-4478 (2009).

4. Nourbakhsh, A. et al. Graphene oxide monolayers as atomically thin seeding layers for atomic layer deposition of metal oxides. *Nanoscale* **7**, 10781-10789 (2015).

5. Sangwan, V.K. et al. Quantitatively Enhanced Reliability and Uniformity of High-κ Dielectrics on Graphene Enabled by Self-Assembled Seeding Layers. *Nano Lett.* **13**, 1162-1167 (2013).

6. Jandhyala, S. et al. Atomic Layer Deposition of Dielectrics on Graphene Using Reversibly Physisorbed Ozone. *ACS Nano* **6**, 2722-2730 (2012).

7. Takahashi, N. & Nagashio, K. Buffer layer engineering on graphene via various oxidation methods for atomic layer deposition. *Appl. Phys. Express* **9**, 125101 (2016).

8. Jeong, S. et al. Thickness scaling of atomic-layer-deposited $HfO_2$ films and their application to wafer-scale graphene tunnelling transistors. *Sci. Rep.* **6** 20907 (2016).

9. Xiao, M., Qiu, C., Zhang, Z. & Peng, L. Atomic-Layer-Deposition Growth of an Ultrathin $HfO_2$ Film on Graphene. *ACS Appl. Mater. Inter.* **9**, 34050-34056 (2017).

10. Radisavljevic, B., Radenovic, A., Brivio, J., Giacometti, V. & Kis, A. Single-layer $MoS_2$ transistors. *Nat. Nanotechnol.* **6**, 147-150 (2011).

11. Wang, J. et al. Integration of High-κ Oxide on $MoS_2$ by Using Ozone Pretreatment for High-Performance $MoS_2$ Top-Gated Transistor with Thickness-Dependent Carrier Scattering Investigation. *Small* **11**, 5932-5938 (2015).

12. Liu, H., Xu, K., Zhang, X. & Ye, P.D. The integration of high-κ dielectric on two-dimensional crystals by atomic layer deposition. *Appl. Phys. Lett.* **100**, 152115 (2012).

13. Yang, J. et al. Improved Growth Behavior of Atomic-Layer-Deposited High-κ Dielectrics on Multilayer $MoS_2$ by Oxygen Plasma Pretreatment. *ACS Appl. Mater. Inter.* **5**, 4739-4744 (2013).

14. Zou, X. et al. Interface Engineering for High-Performance Top-Gated $MoS_2$ Field-Effect Transistors. *Adv. Mater.* **26**, 6255-6261 (2014).

15. Price, K.M., Schauble, K.E., McGuire, F.A., Farmer, D.B. & Franklin, A.D. Uniform Growth of Sub-5-Nanometer High-κ Dielectrics on $MoS_2$ Using Plasma-Enhanced Atomic Layer Deposition.

*ACS Appl. Mater. Inter.* **9**, 23072-23080 (2017).

16. Qian, Q. et al. Improved Gate Dielectric Deposition and Enhanced Electrical Stability for Single-Layer MoS$_2$ MOSFET with an AlN Interfacial Layer. *Sci. Rep.* **6**, 27676 (2016).

17. Park, J.H. et al. Atomic Layer Deposition of Al$_2$O$_3$ on WSe$_2$ Functionalized by Titanyl Phthalocyanine. *ACS Nano* **10**, 6888-6896 (2016).

18. Fang, H. et al. High-performance single layered WSe$_2$ p-FETs with chemically doped contacts. *Nano Lett.* **12**, 3788-3792 (2012).

19. English, Chris D., et al. Approaching ballistic transport in monolayer MoS$_2$ transistors with self-aligned 10 nm top gates. *Inter. Electron Dev. Meet. Tech. Digest* 5.6.1-5.6.4 (2016).

20. Nourbakhsh, A. et al. 15-nm channel length MoS$_2$ FETs with single-and double-gate structures. *Symp. VLSI Tech.* 28-29 (2015).

21. Zhu, Y. et al. Monolayer Molybdenum Disulfide Transistors with Single-Atom-Thick Gates. *Nano Lett.* **18**, 3807-3813 (2018).

22. Yang, L., et al. 10 nm nominal channel length MoS$_2$ FETs with EOT 2.5 nm and 0.52 mA/µm drain current. *73rd Annual Device Research Conference*. IEEE, 237-238(2015).

23. Xie, L. et al. Graphene-Contacted Ultrashort Channel Monolayer MoS$_2$ Transistors. *Adv. Mater.* **29**, 1702522 (2017).

24. Desai, S.B. et al. MoS$_2$ transistors with 1-nanometer gate lengths. *Science* **354**, 99-102 (2016).